\documentclass[twocolumn]{aa}

\usepackage{hyperref}
\usepackage{graphicx}
\usepackage{natbib}
\bibpunct{(}{)}{;}{a}{}{,} 
\usepackage{txfonts}
\usepackage{xcolor}
\usepackage{listings}
\usepackage{acronym}
\usepackage{deluxetable}
\usepackage[outdir=./]{epstopdf}
\usepackage[switch]{lineno}
\usepackage{soul}

\usepackage{hyperref}
\definecolor{myblue}{HTML}{1F77B4}
\definecolor{mygreen}{HTML}{2CA02C}
\definecolor{myred}{HTML}{D62728}
\definecolor{mymagenta}{HTML}{D33682}
\definecolor{codepurple}{HTML}{C42043}

\hypersetup{
bookmarks=true,         
unicode=true,           
colorlinks=true,        
linkcolor=myred,        
citecolor=myblue,       
filecolor=magenta,      
urlcolor=mygreen        
}

\newcommand{\swift}{\textit{Swift}}

\newcommand{\msun}{M$_{\sun}$}

\newcommand{\Msun}{{\ensuremath{\mathrm{M}_{\odot}}}}

\title{A low-energy explosion yields the underluminous Type IIP SN~2020cxd}
\titlerunning{The LL SN ZTF20aapchqy - slow, faint and long-lived}

\author{S. Yang\inst{1}
\and 
J. Sollerman\inst{1}
\and 
N. L. Strotjohann\inst{2}
\and 
S. Schulze\inst{3}
\and 
R. Lunnan\inst{1}
\and 
E. Kool\inst{1}
\and 
C. Fremling\inst{4}
\and 
D. Perley\inst{5}
\and 
E. Ofek\inst{2}
\and 
T. Schweyer\inst{1}
\and 
E.~C. Bellm\inst{6}
\and 
M.~M. Kasliwal\inst{5}
\and 
F.~J. Masci\inst{7}
\and 
M. Rigault\inst{8} 
\and 
Y. Yang\inst{2,9} 
}
\institute{Department of Astronomy, The Oskar Klein Center, Stockholm University, AlbaNova, 10691 Stockholm, Sweden
\and{Department of Particle Physics and Astrophysics, Weizmann Institute of Science, 234 Herzl St, 76100 Rehovot, Israel}
\and{Department of Physics, The Oskar Klein Center, Stockholm University, AlbaNova, 10691 Stockholm, Sweden}
\and{Division of Physics, Mathematics, and Astronomy, California Institute of Technology, Pasadena, CA 91125, USA}
\and{Astrophysics Research Institute, Liverpool John Moores University, IC2, Liverpool Science Park, 146 Brownlow Hill, Liverpool L3 5RF, UK}
\and{DIRAC Institute, Department of Astronomy, University of Washington, 3910 15th Avenue NE, Seattle, WA 98195, USA}
\and{IPAC, California Institute of Technology, 1200 E. California, Blvd, Pasadena, CA 91125, USA}
\and{Univ Lyon, Univ Claude Bernard Lyon 1, CNRS, IP2I Lyon / IN2P3, IMR 5822, F-69622, Villeurbanne, France}
\and{Department of Astronomy, University of California, Berkeley, CA 94720-3411, USA}
}

\begin{document}
\abstract
{We present observations and analysis of SN 2020cxd, a low-luminosity (LL), long-lived Type IIP supernova (SN). 
This object was a clear outlier in the magnitude-limited SN sample recently presented by the Zwicky Transient Facility (ZTF) Bright Transient Survey.}
{We demonstrate that SN 2020cxd is an additional member of the group of LL SNe, 
and discuss the rarity of LL SNe in the context of the ZTF survey, and how further studies of these faintest members of the 
core-collapse (CC) SN family might help understand the underlying initial mass function for stars that explode.} 
{We present optical light curves (LCs) from the ZTF in the $gri$ bands and several epochs of ultra-violet data from the \textit{Neil Gehrels Swift Observatory} as well as a sequence of optical spectra. We construct colour curves, a bolometric LC, compare ejecta-velocity and black-body temperature evolutions for LL SNe, as well as for typical Type II SNe. Furthermore, we adopt a Monte Carlo code that fits semi-analytic models to the LC of SN 2020cxd, which allows the estimation of physical parameters.
Using our late-time nebular spectra, we also compare against SN II spectral synthesis models from the literature to constrain the progenitor properties of SN 2020cxd.}
{The LCs of SN 2020cxd show great similarity with those of LL SNe IIP, in luminosity, timescale and colours.
Also the spectral evolution of SN 2020cxd is that of a Type IIP SN. The spectra show prominent and narrow P-Cygni lines, indicating low expansion velocities. 
This is one of the faintest LL SNe observed, with an absolute plateau magnitude of $M_{r} = -14.5$ mag, and also one with the longest plateau lengths, with a duration of 118 days.
Finally, the velocities measured from the nebular emission lines are among the lowest ever seen in a SN, with intrinsic Full Width at Half Maximum
of 478 km s$^{-1}$.
The underluminous late-time exponential LC tail indicates that the mass of $^{56}$Ni ejected during the explosion is much smaller than the average of normal SNe IIP, we estimate $M_{^{56}Ni}$ = 0.003 \msun. 
The Monte Carlo fitting of the  bolometric LC suggests that the progenitor of SN 2020cxd had a radius of
$R_{0} = 1.3\times10^{13}$ cm, 
kinetic energy of
$E_{kin}=4.3\times10^{50}$ erg, 
and ejecta mass
$M_{\rm{ej}} = 9.5$ \msun.
From the  bolometric LC, we estimate the total radiated energy E$_{\rm{rad}} =  1.52 \times 10^{48}$ erg.
Using our late-time nebular spectra, we compare against SN II spectral
synthesis models to constrain the progenitor Zero-age Main-sequence mass and found it likely to be  $\lesssim 15$ \msun.}
{SN 2020cxd is a LL Type IIP SN. The inferred progenitor parameters and the features observed in the nebular spectrum favour a low-energy, Ni-poor, iron CC SN from a low mass ($\sim12$ \msun)~red  supergiant.}


\maketitle

\section{Introduction}
\label{section:intro}

Stars more massive than about 8 \msun~end their lives with the collapse of their iron core. Type II supernovae (SNe) are the most common among these core-collapse (CC) explosions, and are characterized by the presence of hydrogen in their spectra. 
Type II SNe constitute a diverse class, with light curves (LCs) showing different decline rates across a continuum, and a large range in luminosities, with $V$-band maximum absolute magnitudes ranging from about $-13.5$ to $-19$ mag \citep{Anderson_2014}.
 
Low-luminosity SNe II (LL SNe II)
make up a small part on the faint tail of the detected Type II SN distribution. 
There are only a dozen of such objects presented in the literature (see Table \ref{tab:llsn}), including three recently published examples \citep[][]{Jager2020, Bravo2020, Reguitti2020}.
These are SNe with a faint plateau (SNe IIP) and a late LC tail signalling radioactive powering by a small amount of ejected $^{56}$Ni, typically 
{$\sim0.005$~\msun}
~\citep[][{their fig. 13}]{Spiro14}. 
LL SNe II often also display slow expansion velocities suggesting low explosion energies.
 
It is now well established that 
the progenitors of many SNe~II are red supergiant (RSG) stars \citep{Smartt2009} and it is suspected that the LL SNe IIP originate from stars 
with relatively low zero-age main-sequence (ZAMS) masses  ($\sim8-10$~\msun, \citealt{Pumo16, O'Neill2020}).
However, other studies have suggested the possibility that the progenitors are more massive RSGs with large amounts of fallback material \citep{Zampieri03},
or electron capture supernova (ECSN) explosions of super-asymptotic giant branch (SAGB) stars \citep{Hiramatsu2020}.
There are several methods available to derive information about the exploded star, including identifying SN progenitors in archive images \citep[e.g.,][]{Smartt2009}, light curve modeling  \citep[e.g.,][]{Arnett1989}, and late-time nebular spectral modeling \citep[e.g.,][]{Jerkstrand2012, Jerkstrand2018, Dessart, Lisakov16}. 

For a typical Initial Mass Function (IMF), over 40\% of the potential CC SN progenitors reside in the $8-12$ \Msun~
range and 25\% in the $8-10$ \Msun~range \citep{Sukhbold2016}.
The fact that the LL SNe are rarely detected and under-explored is of course due to the many observational biases challenging the discovery and follow-up of such faint transients. Most of them have been discovered in targeted searches of relatively nearby galaxies.
However, with the ongoing revolution in transient science, with systematic non-targeted surveys uncovering thousands of transients, this is beginning to change. 
In this paper we use data from one of the ongoing surveys, the Zwicky Transient Facility \citep[ZTF;][]{Bellm_2018, Graham_2019}.
In particular, \cite{Fremling2020} introduced the ZTF Bright Transient Survey (BTS), which provides a large and purely magnitude-limited sample of extragalactic transients in the northern sky, suitable for detailed statistical and demographic analysis.  
\cite{Perley2020} presented early results of the BTS, which is almost spectroscopically complete down to a target magnitude of 18.5. In particular, they presented a CC SN luminosity function, in which a significant fraction of the CC SNe is very dim. 
They state that this is in agreement with works arguing that the ``SN rate problem'' (\citealt{Horiuchi2011}) can be resolved using galaxy-untargeted surveys and including the faint end of the SN  luminosity function.
 
Similar efforts to construct an unbiased view of the Type II SN luminosity function comes from the parallel ZTF volume-limited survey, the ZTF Census of the Local Universe (CLU) experiment, which extends the classification threshold to $m\lesssim20$ mag for transients occurring in known galaxies within $D<200$ Mpc \citep{De2020}. These results also imply that the lowest luminosity events are more common than previously appreciated (Tzanidakis et al. in prep.). The ultimate aim of such large efforts is to be able to connect SN  rates with star-formation rates and to couple the stellar evolution 
IMF to the known SN  populations.

Out of the 171 Type II SNe presented by \cite{Perley2020} one object clearly stands out as both longer duration and considerably fainter than the rest of the population, SN 2020cxd \citep[a.k.a. ZTF20aapchqy,][their fig. 7]{Perley2020}. 
The authors declare that this is "almost certainly the explosion of a massive star".
In this paper we take the opportunity to present  SN\,2020cxd in more detail. 
We argue that this transient is not only interesting as a single object given the rarity of LL SNe in the literature, but that this event is 
important in context with the overall population of SNe II provided by the strict and clear criteria of the highly complete BTS sample.

The paper is structured as follows.
In Sect.~\ref{sec:data}, we outline the observations and corresponding data reductions, including Sect.~\ref{sec:detection} where we present the discovery and classification of SN\,2020cxd.
The ground-based optical SN imaging observations and data reductions are presented in Sect.~\ref{sec:optical}, and
in Sect.~\ref{sec:space} we describe the \textit{Swift} observations. A
search for pre-explosion outbursts is done in
Sect.~\ref{sec:prediscovery}, the optical spectroscopic follow-up campaign is presented in Sect.~\ref{sec:opticalspectra}, and a discussion of the host galaxy is provided in Sect.~\ref{sec:host}.
An analysis and discussion of the results 
is given in Sect.~\ref{sec:discussion} 
and we provide a summary in Sect.~\ref{sec:summary}.
Some of the more technical aspects of the analysis are presented in the Appendices.

\section{Observations and Data reduction}
\label{sec:data}

\subsection{Discovery and classification}
\label{sec:detection}

SN\,2020cxd was first discovered with the Palomar Schmidt 48-inch (P48) Samuel Oschin telescope on February 19, 2020 ($\mathrm{JD_{discovery}}=2458899.0306$), as part of the ZTF survey. 
The first ZTF detection was made in the $r$ band,
with a host-subtracted magnitude
of $17.69\pm0.05$~mag, at the J2000 coordinates
$\alpha=17^{h}26^{m}29.26^{s}$, $\delta=+71\degr05\arcmin38.6\arcsec$. 
The first $g$-band detection came in 31 minutes later at $17.63\pm0.04$.
An on-duty astronomer (JS) immediately triggered follow-up observations, and several telescopes started observing shortly thereafter. 
The discovery was reported to the Transient Name Server 
(TNS\footnote{\href{https://www.wis-tns.org/}{https://www.wis-tns.org/}}) 
by \cite{2020TNSTR.555....1N} with a note that the last non-detection was three days earlier on February 16,
with a global limit of 20.4 in the $g$ band. 
We can therefore constrain the epoch of explosion for this SN with good precision. 
In this paper we will adopt an explosion date of
$\mathrm{JD_{explosion}}=2458897.5301$ with an uncertainty of 
$\pm1.5$ days, as given by the epoch half way between discovery and last 
non-detection\footnote{Following the methods of \cite{Bruch_2020}, we found there were not enough early phase data in either band to perform a power-law fit, and we thus set the explosion epoch as the mean of the first detection and the last non-detection.}. 

\begin{figure*}
\centering
    \includegraphics[width=0.5\textwidth]{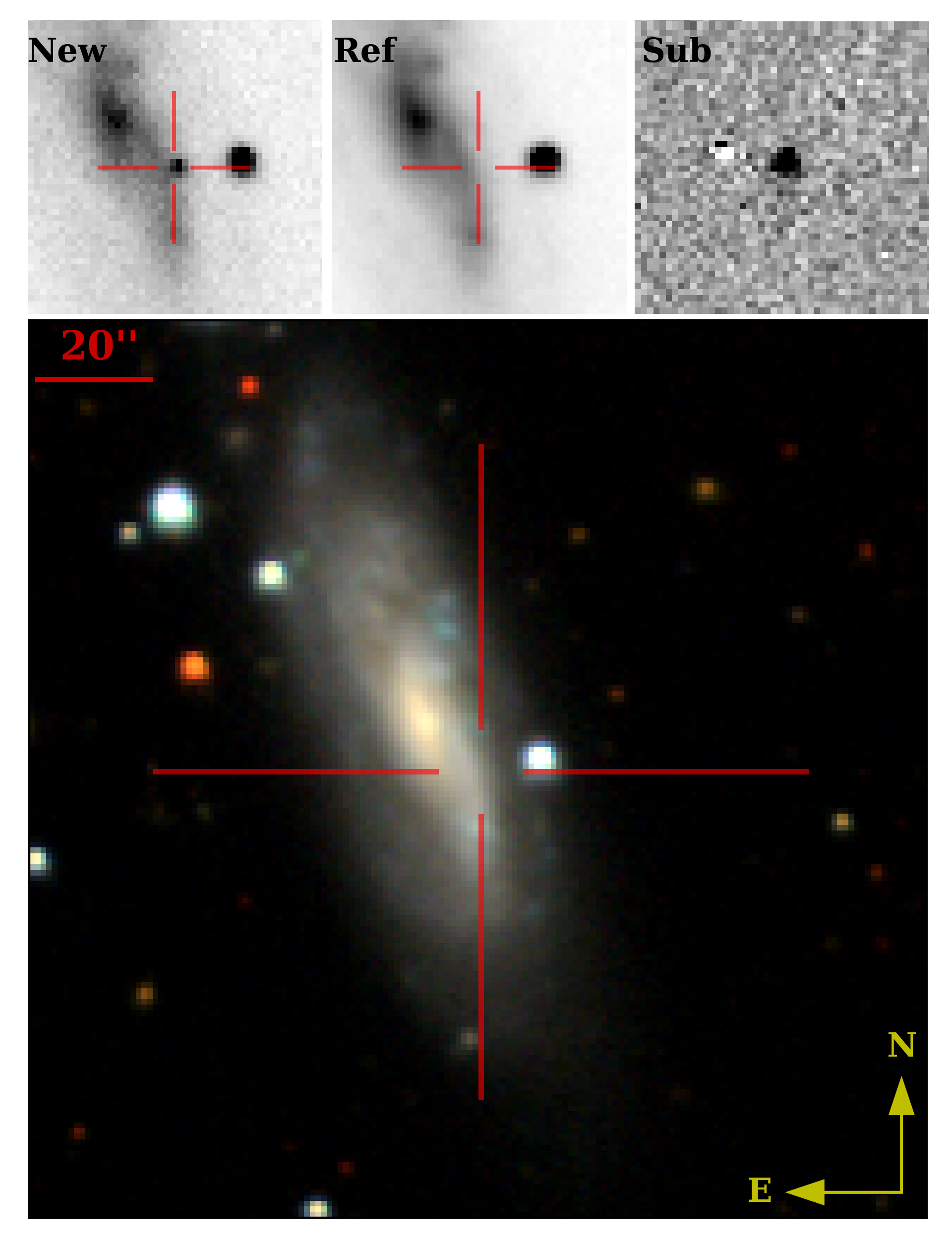}
    \caption{SN\,2020cxd in the nearby galaxy NGC 6395.
    The $g$-band image subtraction is shown in the top panels (subtraction to the right), 
    with the SN image observed 
    on 2020 April 24, 64.9 days after the first ZTF detection (to the left).
    The bottom panel shows a $gri$-colour composite image of the host galaxy and its environment.
    It was composed of ZTF $g-$, $r-$ and $i-$band pre-explosion images of the field.} 
    \label{fig:det}
\end{figure*}

SN\,2020cxd is positioned in the nearby spiral galaxy NGC 6395
which has a measured redshift of $z=0.003883$.
According to the NASA/IPAC Extragalactic Database (NED)\footnote{\href{https://ned.ipac.caltech.edu}{https://ned.ipac.caltech.edu}} catalog, the peculiar motion corrected distance for a 
standard cosmology ($\Omega_M=0.3$, $\Omega_\Lambda=0.7$, and $h=0.7$) is 23 Mpc, whereas the most recent Tully-Fisher measurement reported on the same site is
20 Mpc. In this paper we will adopt a distance modulus of 31.7$\pm$0.3 mag
($22\pm3$ Mpc).
We also obtain the amount of Galactic extinction along the line of sight using the 
NED extinction tool\footnote{\url{https://ned.ipac.caltech.edu/extinction_calculator}} (based on the dust map of \citealp{Schlafly11}) and adopt
E($B-V$) = 0.035 mag. 
The SN together with the host galaxy and the field of view is shown in Fig.~\ref{fig:det}.

SN\,2020cxd was classified as a Type II SN
\citep{2020TNSCR1503....1P,2020TNSCR.576....1P}
based on a spectrum obtained on February 20, 2020, 17 hours past discovery,
with the Liverpool telescope (LT) equipped with the SPectrograph for the Rapid Acquisition of Transients (SPRAT). 
That spectrum revealed a blue continuum with hydrogen P-Cygni features (broad H$\alpha$ and H$\beta$).
The classification was consolidated with a spectrum taken 4 hours thereafter with the 
Palomar 60-inch telescope (P60; \citealp{2006PASP..118.1396C}) equipped with the Spectral Energy Distribution Machine (SEDM;
\citealp{SEDM}).

\subsection{Optical photometry}
\label{sec:optical}

Following the discovery, we obtained regular follow-up photometry
during the 100+ days long plateau phase in $g$, $r$ and $i$ bands
with the ZTF camera \citep{dekany2020} on the P48. Early LT photometry in 
$ugriz$ was also obtained at one epoch to measure the colours, and a campaign with the {\tt Swift} observatory was launched (Sect.~\ref{sec:space}).
Later on, after the drop from the plateau, we also obtained a few epochs
of photometry in $gri$ with the SEDM on the P60, with the LT telescope and with the Nordic Optical telescope (NOT) using
the Alhambra Faint Object Spectrograph (ALFOSC). 
The photometric magnitudes of SN\,2020cxd are listed in Table \ref{tab:ground}.

\begin{figure*}
\centering
    \includegraphics[width=0.8\textwidth]{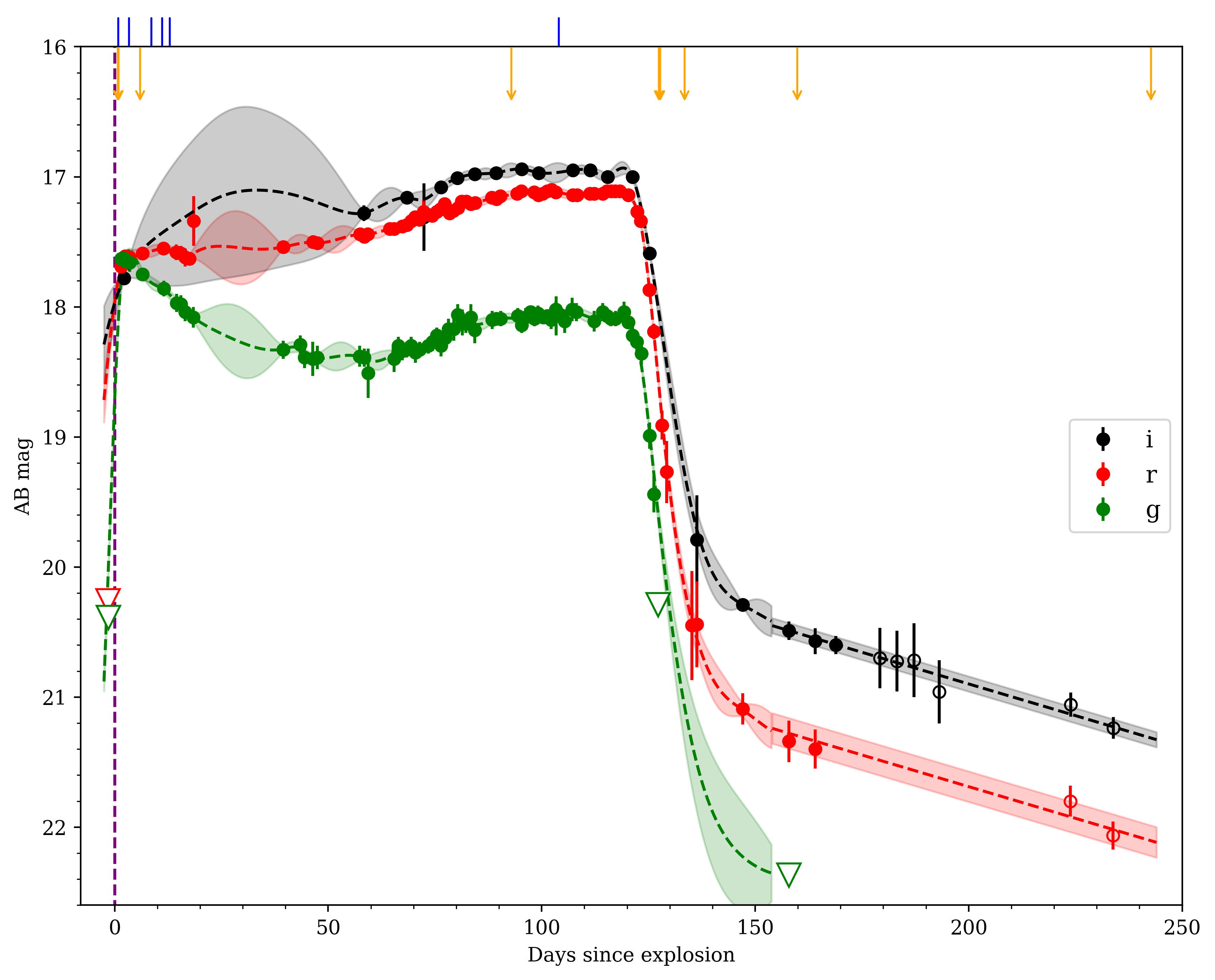}
    \caption{ 
    Light curves of SN\,2020cxd in the $g$ (green symbols), $r$ (red) and $i$ (black) bands. These are host subtracted magnitudes from the P48, P60 and LT, as well as two epochs from the NOT on the late tail. 
    Forced photometry obtained using SNOoPY is plotted as open circles. All are observed (AB) magnitudes in the observer frame in days since the explosion. 
    Relevant upper limits are displayed as inverted triangles, and constrain the explosion epoch of the SN (the purple vertical dashed line), as well as the $g$-band tail.
    The dashed lines with error regions are Gaussian Process estimates of the interpolated/extrapolated LC. After 150 days we instead show linear fits.
    The yellow downward pointing arrows on top indicate the epochs of spectroscopy, the blue vertical bars display the epochs with {\it Swift} observations.
    }
    \label{fig:lc}
\end{figure*}

The LCs from the P48 come from the ZTF pipeline \citep{2019PASP..131a8003M}.
Photometry from the P60 were produced with the image-subtraction pipeline described in \cite{fremling16}, 
with template images from the Sloan Digital Sky Survey (SDSS; \citealp{ahn14}).
This pipeline produces point spread function (PSF) magnitudes, calibrated against SDSS stars in the field. 
For the late NOT images,  
template subtraction was performed with {\tt hotpants}\footnote{\url{ http://www.astro.washington.edu/users/becker/v2.0/hotpants.html}}, using archival SDSS images.
The magnitudes of the transient were then measured using SNOoPY\footnote{SNOoPy is a package for SN photometry using PSF fitting and template subtraction developed by E. Cappellaro. A package description can be found at \url{http://sngroup.oapd.inaf.it/snoopy.html}.} and calibrated against SDSS stars in the field.
All magnitudes are reported in the AB system.
The reddening corrections are applied using the \cite{1989ApJ...345..245C} extinction law with $R_V=3.1$. 
We do not correct for host galaxy extinction, since there is no sign of narrow
\ion{Na}{i d} absorption lines in our spectra. This is basically an assumption, and we discuss some of the implications of this towards the end of the paper.
The $gri$ LCs are shown in Fig.~\ref{fig:lc}.

We used a Gaussian Process (GP) algorithm\footnote{\href{https://george.readthedocs.io}{https://george.readthedocs.io}}
to interpolate the photometric measurements and found\footnote{via $scipy.find\_peaks$}
that the peak happened at m$^{\rm{peak}}_{r} = 17.14 \pm 0.01$ after t$^{\rm{rise}}_{r} = 109.50 \pm 1.60$ rest frame days past explosion. 
In the $i$ band, the photometric behavior followed the same trend and reached a maximum at
m$^{\rm{peak}}_{i} = 17.00$ after 
t$^{\rm{rise}}_{i} = 113.46$ rest frame days.
The $g$-band light curve actually peaks at the first detection epoch. It slowly declines before it flattens out on a plateau which reached a maximum at m$_{g} = 18.08$ after 115.26 days.
Thereafter the SN declined fast, and soon became too faint for detection with the P48.
We then activated the LT and NOT telescopes.
The $r$- and $i$-band observations on the tail suggest a linear decline.
We performed a linear fit after $\sim$ 150 days, and found that the best fit slopes for the $r$ and $i$ band are $(8.86\pm1.33)\times10^{-3}$ and $(9.99\pm0.79)\times10^{-3}$
mag days$^{-1}$, which are consistent with the radioactive cobalt decay slope ($\lambda_{Co}=1/111.3\times2.5/ln(10)\sim9.76\times10^{-3}$ mag~days$^{-1}$).
For the $g$-band images on the tail, we found no apparent flux in the LT and NOT residual images down to 2.5 sigma (upper limits are also presented in
Table~\ref{tab:ground}).

\subsection{Swift-observations: UVOT photometry\label{sec:space}}

A series of ultraviolet (UV) and optical photometry observations were obtained with the UV/Optical Telescope onboard the Neil Gehrels \swift\ observatory (UVOT; \citealp{gehrels2004}; \citealp{2005SSRv..120...95R})
between February 20 and June 2, 2020.
Our first {\it Swift}/UVOT observation was performed on February 20, 2020, ($\mathrm{JD}=2458899.8624$), 0.83 days past discovery (2.3 days past estimated explosion). The SN was detected in all bands. 
The brightness in the UVOT filters was measured with UVOT-specific tools in 
HEAsoft\footnote{\href{https://heasarc.gsfc.nasa.gov/docs/software/heasoft/}{https://heasarc.gsfc.nasa.gov/docs/software/heasoft/} version 6.26.1.}.

Source counts were extracted from the images using a circular region with a radius of $3''$. The background was estimated using a circular region with a radius of $39''$. The count rates were obtained from the images using the \swift\ tool {\tt uvotsource} and converted to magnitudes using the UVOT photometric zero points \citep{Breeveld2011a} and the new calibration data from September 2020\footnote{\href{https://heasarc.gsfc.nasa.gov/docs/heasarc/caldb/swift}{https://heasarc.gsfc.nasa.gov/docs/heasarc/caldb/swift}}. We obtained a final UVOT epoch in August 2020, after the SN had faded, to remove the host contribution from the transient light curve. 
A log of the Swift observations is provided in Table~\ref{tab:swift}. This includes six epochs in total, and these epochs are also indicated in Fig.~\ref{fig:lc}.  Such a dense and early UV coverage is very rare for LL SNe, and is important for constructing the bolometric light curve needed to model the SN in Sect.~\ref{sec:bl}, see also Sect.~\ref{sec:bb}.

\subsection{Pre-discovery imaging}
\label{sec:prediscovery}

Since the SN occurred in a very nearby galaxy we 
decided to search for pre-explosion activity of the progenitor star. For this purpose we downloaded IPAC difference images\footnote{\url{https://irsa.ipac.caltech.edu/applications/ztf/}}, performed forced photometry at the SN position \citep{yuhan2019}, and applied quality cuts and searched for pre-explosion detections as described by \citet{Nora_2020}. 

\begin{figure*}
\centering
    \includegraphics[width=0.8\textwidth]{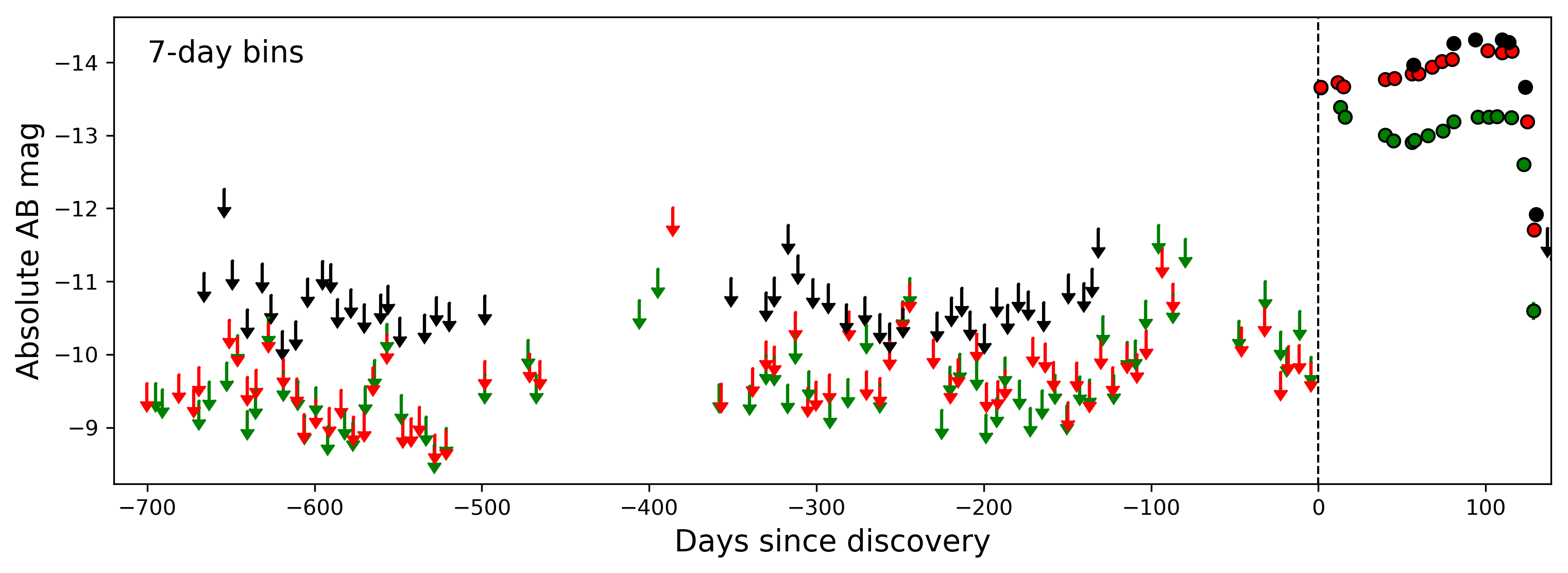}
    \caption{ZTF P48 pre-explosion images for SN\,2020cxd reveal no precursors in $g$ (green symbols), $r$ (red) or $i$ (black) bands. 
    Filled data points are $\gtrsim 5 \sigma$ detections, whereas arrows show   $5 \sigma$ upper limits. 
    The data are binned in 7-day bins.
    }
    \label{fig:precursor}
\end{figure*}

ZTF started to monitor the position of SN\,2020cxd 2.5 years before the discovery and we searched for pre-explosion outbursts in 999 observations collected on 307 different nights that were obtained in the $g$, $r$ and $i$ bands. No outbursts are detected above the $5\sigma$ threshold when searching unbinned or binned (1 to 90-day long bins) LCs. Figure~\ref{fig:precursor} shows the absolute magnitude LC in 7-day-long bins, where filled data points are significant at the $5\sigma$ level while arrows indicate $5\sigma$ upper limits. The $g$-band ($r$-band) observations are available in 69 (68) out of 132 weeks before the SN discovery and we can exclude flares brighter than an absolute magnitude of $-10$ in 46 weeks (41 weeks), i.e. 34\% (31\%) of the time.

\subsection{Optical spectroscopy}
\label{sec:opticalspectra}

\begin{figure*}
\centering
    \includegraphics[width=\textwidth]{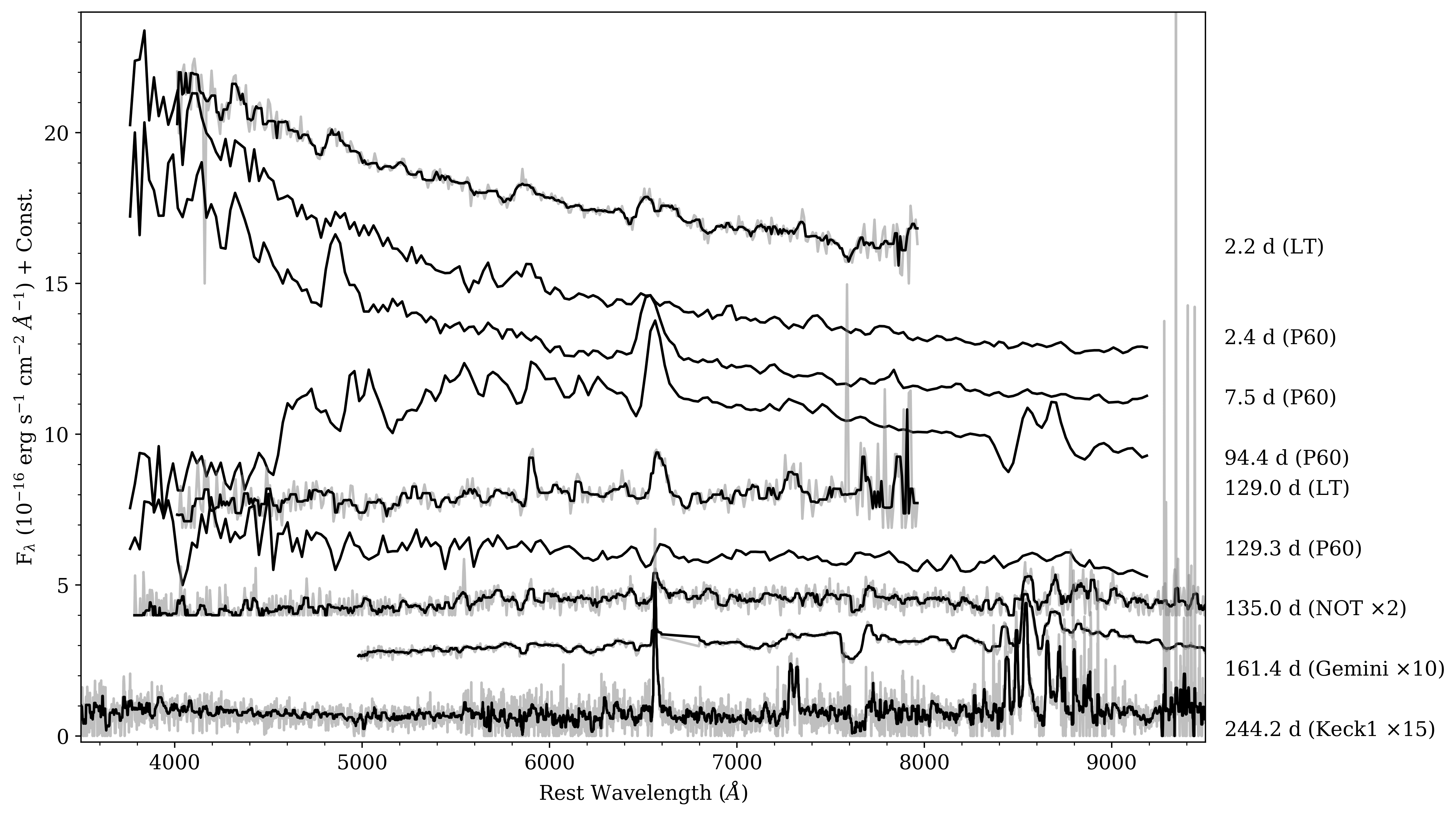}
    \caption{Sequence of optical spectra for SN\,2020cxd, redshift- and reddening-corrected. The complete log of spectra is provided in Table~\ref{tab:spec}. The epoch (rest frame days since explosion) of the spectra is provided to the right. The spectra are shifted in flux by a constant for clarity. Late time spectra are scaled for comparison. The scaling factors are also listed to the right. 
    Some spectra are also binned (shown as black lines), with a window size of 5 $\AA$ for LT, and 10 $\AA$ for NOT/Gemini/Keck, and their original spectra are shown in shaded grey.
    }
    \label{fig:spectra}
\end{figure*}

Spectroscopic follow-up was conducted with SEDM 
mounted on the P60 and with the LT SPRAT spectrograph on La Palma.
Further spectra were obtained with the 
NOT using ALFOSC 
as well as one epoch with Gemini North and GMOS and a final spectrum with Keck and LRIS.
A log of the 9 obtained spectra is provided in
Table~\ref{tab:spec}.
SEDM spectra were reduced using the pipeline described by
\citet{RigaultPySEDM} and the LT and NOT spectra were reduced using standard pipelines.
The spectra were finally absolute calibrated against the $r$-band photometry using the GP interpolated
magnitudes and then corrected for Milky Way (MW) extinction.
All spectral data and corresponding information is made available via WISeREP\footnote{\href{https://wiserep.weizmann.ac.il}{https://wiserep.weizmann.ac.il}} \citep{wiserep}.
We present the sequence of spectra in Fig.~\ref{fig:spectra}, and the epochs of spectroscopy are also highlighted in Fig.~\ref{fig:lc}. Overall, we managed to acquire spectra over the full relevant range of the SN evolution, but it should also be noted that the photospheric part of the evolution is mainly covered by robotic telescopes providing limited resolution and signal.

\subsection{Host galaxy}
\label{sec:host}

NGC 6395 is a nearby grand 
Scd spiral galaxy in the Draco constellation, which has not been reported to host any SN before. Simbad reports an AB magnitude of 13 in the $r$ band, but there is no SDSS spectrum of this galaxy and we have not been able to find any estimates of the metallicity of NGC 6395. 
We therefore made use of some of the spectroscopic observations of the SN to also measure some of the stronger emission lines from the nearby regions as a way to probe the metallicity of the gas close to the site of the explosion.
The late Keck spectrum was taken with the slit oriented through the galaxy, and we extract the spectrum of an H~II region located 8\arcsec north of the SN  to probe the gas-phase metallicity there. We discuss the results in Sect.~\ref{sec:host_analysis}.

\section{Analysis and Discussion}
\label{sec:discussion}

\subsection{Light curves}
\label{sec:lc}

The $g$-, $r$- and $i$-band LCs of our SN are displayed in
Fig.~\ref{fig:lc}.
In the figure we have also
included the most restrictive upper limits as triangles ($2.5\sigma$),
while the arrows on top of the figure illustrate epochs of
spectroscopy. The GP interpolation (at early phases) and linear fits (during nebular phases) are also shown.

\begin{figure*}
\centering
    \includegraphics[width=\textwidth]{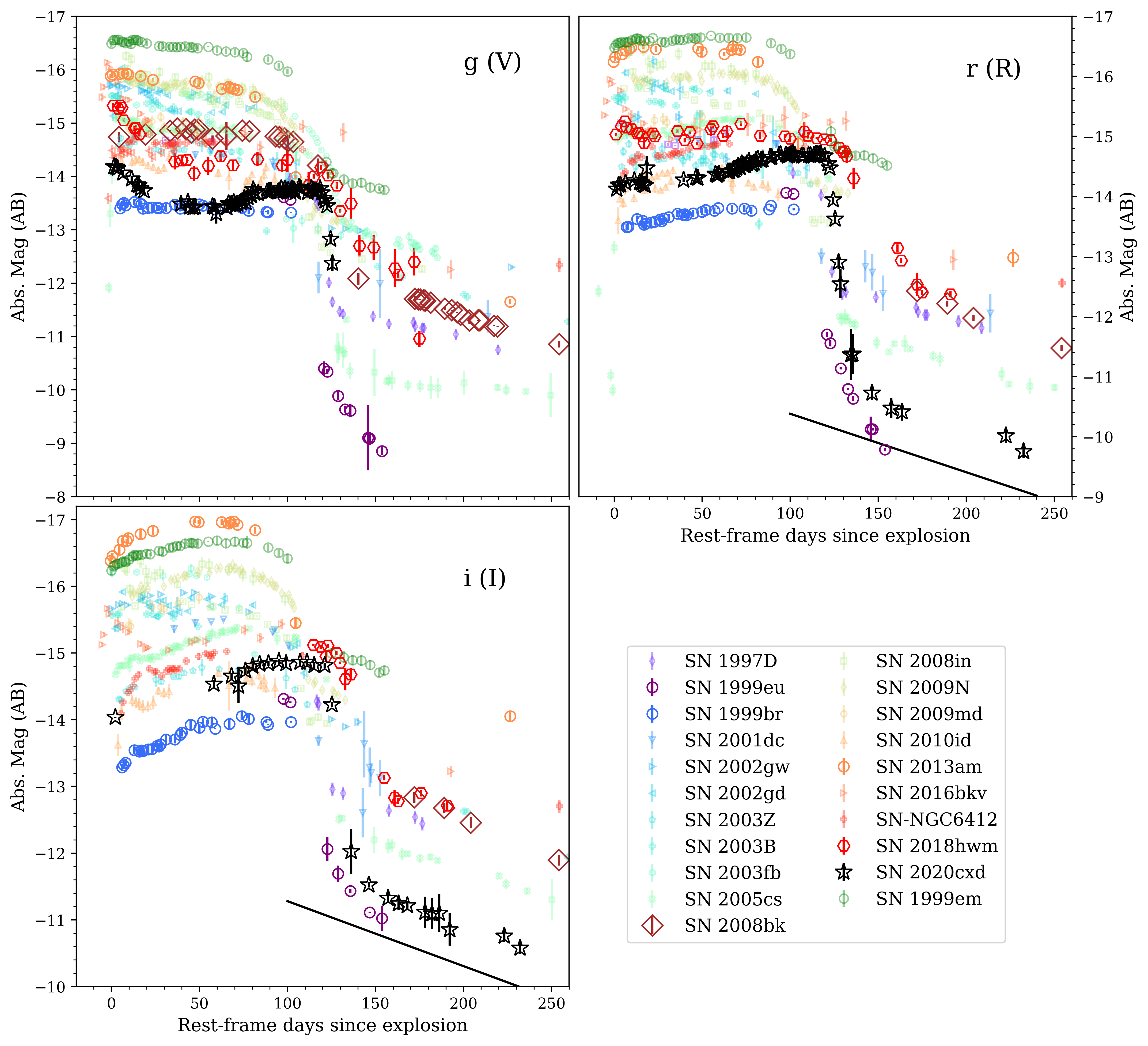}
    \caption{Absolute magnitudes of SN\,2020cxd (black stars) together with the LCs of other LL SNe IIP, as well as for the normal Type II SN 1999em (green circles).
    We highlight SN 2013am (orange circles) which is the brightest within the LL SN II sample, and SNe 1999eu (purple circles) and 1999br (blue circles) which are the faintest. 
    We also use a larger marker for SN 2018hwm (red hexagons), which was also monitored by ZTF in $gri$ photometry.
    The other SNe are displayed with smaller and fainter markers.
    The LCs are plotted versus rest frame days past explosion.
    The radioactive cobalt decay rate is shown as a black line in the $r$- and $i$-band panels for comparison.
    The photometric data were obtained via the \href{https://sne.space/}{Open Supernova Catalog}.
    Data reference for SN 1999em is \cite{Leonard2002}, and for the LL SNe II these are listed in Table~\ref{tab:llsn}. }
    \label{fig:lc+iptf}
\end{figure*}

It is clear that SN\,2020cxd must have risen very fast in the first few days.
The rise is 2.57 magnitudes ($r$ band) in 3 days or less, with the slope thus larger than 0.85 mag per day.
This fast rise of SN\,2020cxd is comparable to that of the LL SN 2005cs (0.90 mag/day in the $R$ band, 
see Fig.~\ref{fig:lc+iptf} 
and \citealt[][their fig. 4]{Pastorello_2009}).

After that initial rise follows a plateau phase of 100+ days, which firmly establishes SN\,2020cxd as a Type IIP SN. The plateau is, however, not completely flat. In the $g$ band there is an initial decline which levels off and then very slightly rises until the end of this phase. The $r$-band LC shows an initial undulation, but is then slowly rising by about 0.5 mag in 100 days, and the $i$ band is following in that rise. The LC is well sampled with 58 observations ($r$ band) over the plateau that allow us to discern such unusual LC structures. 

The drop from the plateau is sharp and fast.
The $g$-band LC declines by 4.98 mag in only 28 days.
The drop in the $r$ and $i$ bands are slightly shallower, making SN\,2020cxd redder at the very late-time phases, but the better sampled $r$-band LC also sharply plummets by 1.9 mag in only 6 days.

When comparing to the large compilation of Type II LCs from \cite{Anderson_2014} and using their nomenclature\footnote{With SN\,2020cxd in the $g$ band, whereas their sample was in $V$ band.},
we find that the Optically Thick Phase duration (OPTd) of SN\,2020cxd is $118.3 \pm 3.0$ d\footnote{As done by \cite{Anderson_2014}, we fit $g$-band data using a $\chi^2$ minimizing procedure with a composed function of Gaussian, Fermi Dirac and straight line, following \cite{Olivares_E__2010}, from the estimated time of explosion.},
which is one of the longest plateau lengths compared to the Anderson sample,
and the decline rate during plateau phase ($s_{2}$) is $-0.73$ mag per 100 days, which is also very rare\footnote{As shown in \citet[][their fig. 2]{Anderson_2014}, the mean of the $s_2$ distribution is 1.27 mag per 100 days with a variance of 0.93 mag. The $s_2$ slope of SN\,2020cxd is 2.2 $\sigma$ off the distribution. However, there are a few SNe in their sample with similarly negative $s_2$. We have been made aware, however, that such a negative plateau slope might be more common among the LL SNe II, see for example SN\,1999br \citep{Pastorello04}, SN\,2009N \citep{takats2014} and SNe 2013K and 2013am \citep{Tomasella2018}. }.
Most of the SNe IIP have instead a clearly positive $s_2$ slope, i.e the plateau is slowly declining in luminosity, and 
SN\,2020cxd is thus an unusually well sampled example of a brightening plateau.

In Fig.~\ref{fig:lc+iptf} we show the $r$/$R$-, $g$/$V$- and $i$/$I$-band LCs in
absolute magnitudes together with the LCs of several other LL SNe II (see Table~\ref{tab:llsn} for detailed information). 
Since most of the LL SNe II in the literature were observed in the Johnson photometric system, 
we compare LCs between $r/R$, $g/V$, and $i/I$, correspondingly.
They show similarity with the canonical Type II plateau SN 1999em (green circles).
The magnitudes in Fig.~\ref{fig:lc+iptf} are in the AB 
system\footnote{The Vega/AB magnitude conversion follows \cite{Blanton2007}.}
and have been
corrected for distance modulus, MW extinction and host extinction if any, and are plotted
versus rest frame days past estimated explosion epoch (see Table~\ref{tab:llsn}).

The figure demonstrates that SN\,2020cxd is a clear member of this population of LL SNe, and in many respects a rather typical representative example for this class. The better photometric sampling compared to most other objects allows to clearly see the on-plateau evolution. The well determined explosion epoch and the sharp drop from the plateau also allows a more precise measure of the plateau duration (OPTd).

\begin{figure*}
\centering
    \includegraphics[width=0.8\textwidth]{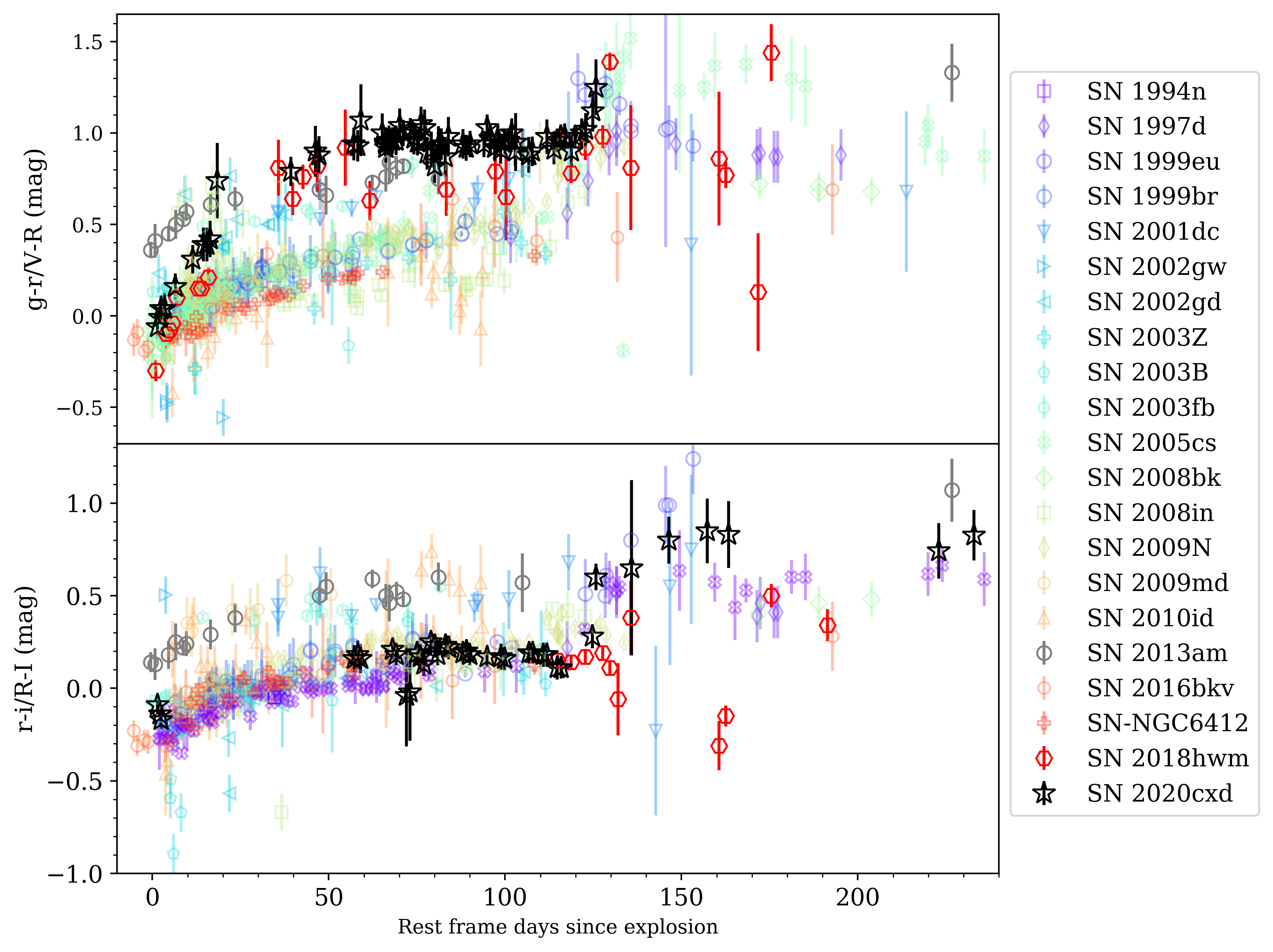}
    \caption{Colour evolution of SN\,2020cxd (black stars) shown in $g-r$ (upper panel) and $r-i$ (lower panel). The colours have been corrected for MW extinction. For comparison we have also plotted colours for several other LL SNe II. 
    Most of these are $V-R$ and $R-I$.
    For references see Table~\ref{tab:llsn}.}
    \label{fig:color}
\end{figure*}

The colour evolution of SN\,2020cxd and the other LL SNe is shown in Fig.~\ref{fig:color}.
We plot ($g-r$)/($V-R$) in the upper panel and ($r-i$)/($R-I$) in the lower panel, both corrected for MW extinction. In doing this, no interpolation was used. Given the excellent LC sampling we used only data where the pass-band magnitudes were closer in time than 0.1 days.
Comparison is made with the colour evolution for other LL SNe, which are known to typically be redder than ordinary SNe II. 
Compared to this sample, our SN is similar, which implies that the amount of additional host extinction is indeed likely small compared to the rest of this population, in accordance with our assumption of negligible host extinction.  Our SN shows normal $g-r$ colors at early phases, and evolves towards the redder part of the sample population after about a month. The $r-i$ stays in the lower part of the LL SN II color space over the entire plateau phase.


\subsection{Spectra}
\label{sec:spectra}

The log of the spectroscopic observations is provided in 
Table~\ref{tab:spec} and the sequence of spectra is shown in Fig.~\ref{fig:spectra}.
Overall, the spectra of SN\,2020cxd are typical for a Type II SN. In this section, we compare the spectra of SN\,2020cxd with spectra from other LL Type IIP SNe at different phases. 

\begin{figure*}
\centering
    \includegraphics[width=\textwidth]{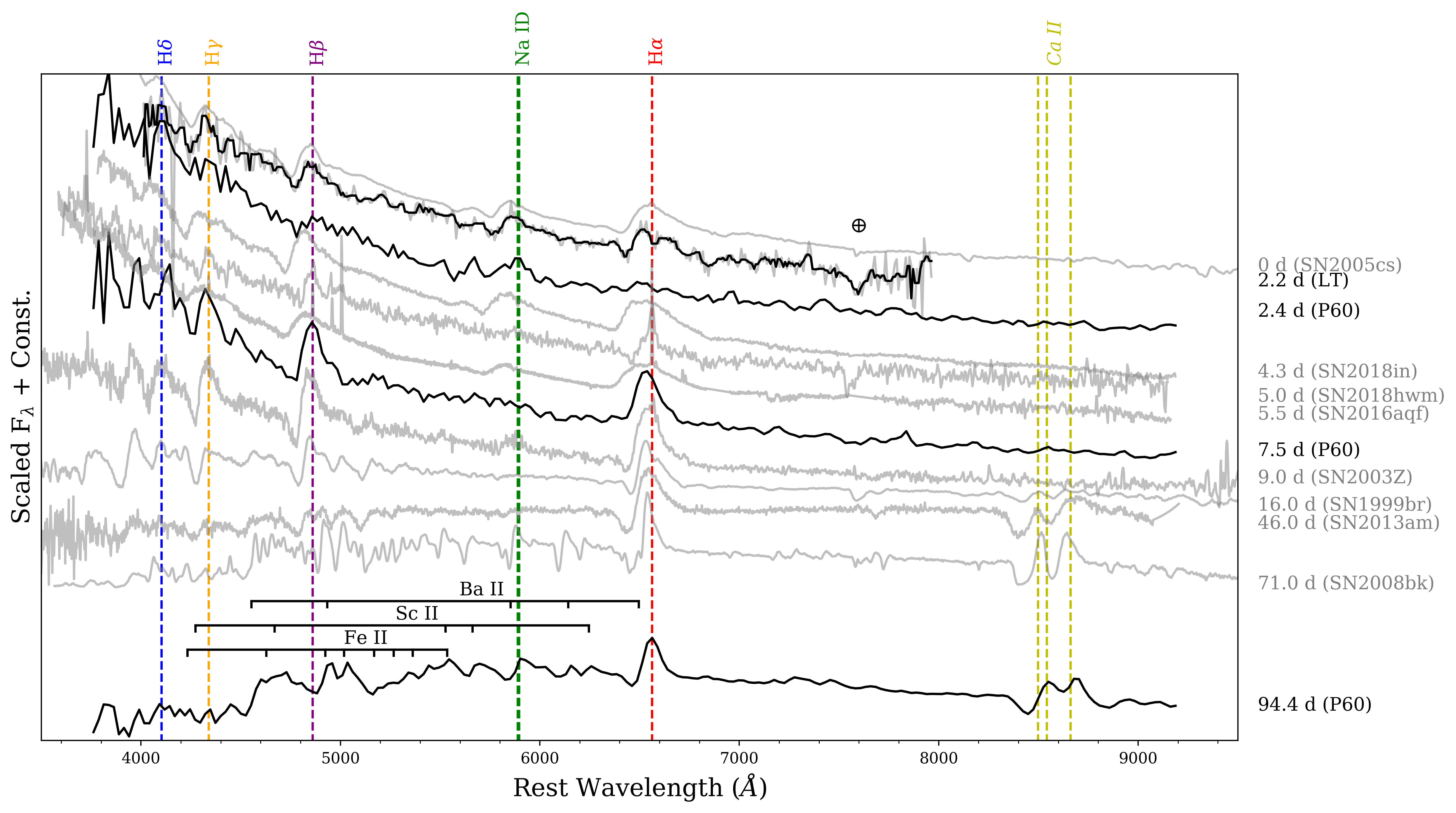}
    \caption{Sequence of early photospheric spectra for SN\,2020cxd, compared with spectra of other LL Type IIP SNe in grey. The epoch of the spectrum is provided to the right.
    Some identified lines (H, Na I D and Ca NIR triplet) are marked as vertical dashed lines.
    We also mark several metal lines as short vertical lines, which might be present in the last P60 spectrum.
    }
    \label{fig:spectra_early}
\end{figure*}

Spectral comparisons between SN\,2020cxd and other LL SNe IIP at photospheric phases are provided in Fig.~\ref{fig:spectra_early}.
All spectra have been corrected for redshift and galactic reddening.
Within $\sim$10 days after explosion, all of the comparison SNe show very blue continuum, with dominant line features of Balmer lines and 
Na I D\footnote{Could also be blended with \ion{He}{I}~$\lambda$5876, \citep[see e.g.,][their fig. 5]{Pastorello_2006}.}.

As SN\,2020cxd enters the plateau phase, the spectrum becomes redder as the photospheric temperature decreases. 
At +93 days, the \ion{Ca}{II} near-infrared (NIR) triplet of SN\,2020cxd is well developed, and we can also discern several metal lines, e.g. \ion{Fe}{II}, \ion{Sc}{II}, and \ion{Ba}{II} (Fig.~\ref{fig:spectra_early}).

\begin{figure*}
\centering
    \includegraphics[width=\textwidth]{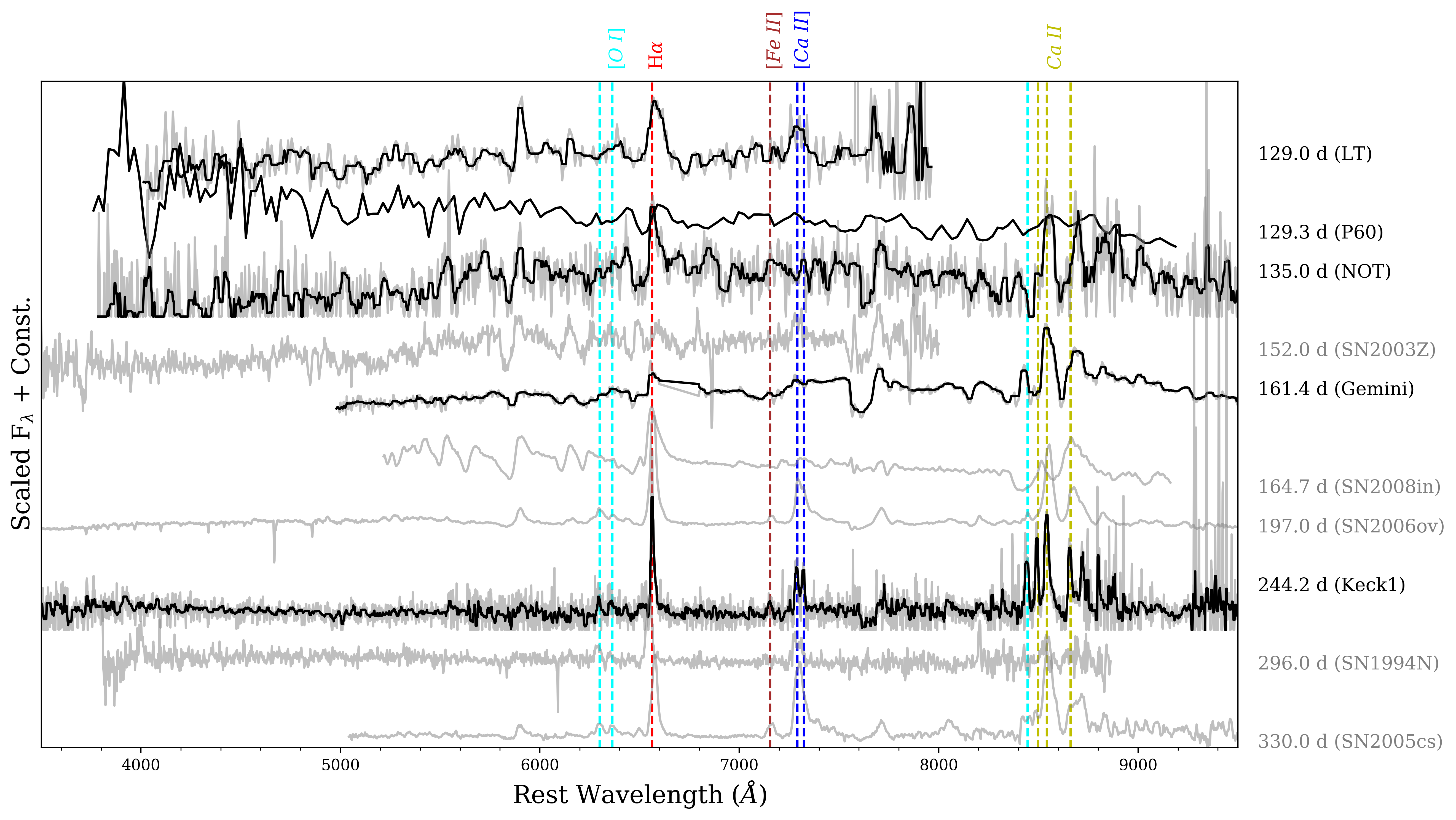}
    \caption{Sequence of nebular spectra for SN\,2020cxd, compared with spectra of other LL Type IIP SNe in grey. The epoch of the spectrum is provided to the right.
    Some identified lines, e.g., H$\alpha$, [\ion{Ca}{II}]~$\lambda\lambda$7291, 7323, \ion{Ca}{II} NIR triplet, [\ion{O}{I}]~$\lambda\lambda$6300, 6364 and $\lambda$8446, are marked as vertical dashed lines.}
    \label{fig:spectra_nebula}
\end{figure*}

From $\sim$120 days, SN\,2020cxd enters the optically thin phase, and the luminosity decreases rapidly. 
The spectra displayed in Fig.~\ref{fig:spectra} are all absolute calibrated against $r$-band photometry as interpolated from the LC, and the last three nebular spectra have been scaled by factors of 2, 10, and 15 respectively, for easier comparison.
Figure~\ref{fig:spectra_nebula} shows the scaled late-time spectra compared to other LL SNe II at similar phases.
From 160 days, the spectra are dominated by H$\alpha$, [\ion{Ca}{II}]~$\lambda\lambda$7291, 7323 and the \ion{Ca}{II} NIR triplet. 

During the photospheric phase, we measured the velocities for SN\,2020cxd using {\tt iraf} \footnote{{IRAF} is distributed by the National Optical Astronomy Observatories, which are operated by the Association of Universities for Research in Astronomy, Inc., under cooperative agreement with the National Science Foundation.} to fit a Gaussian to the minimum of the absorption lines of the corresponding P-Cygni profiles.  
Velocities and their uncertainties were estimated by a random sampling on the Gaussian fits to the minimum, by shifting the anchor continuum points a 1000 times within a region of $\pm 5 \AA$, using {\tt astropy.modeling.Gaussian1D} for the iteration. Fits with bad chi-squared values were excluded.

\begin{figure*}
\centering
    \includegraphics[width=0.8\textwidth]{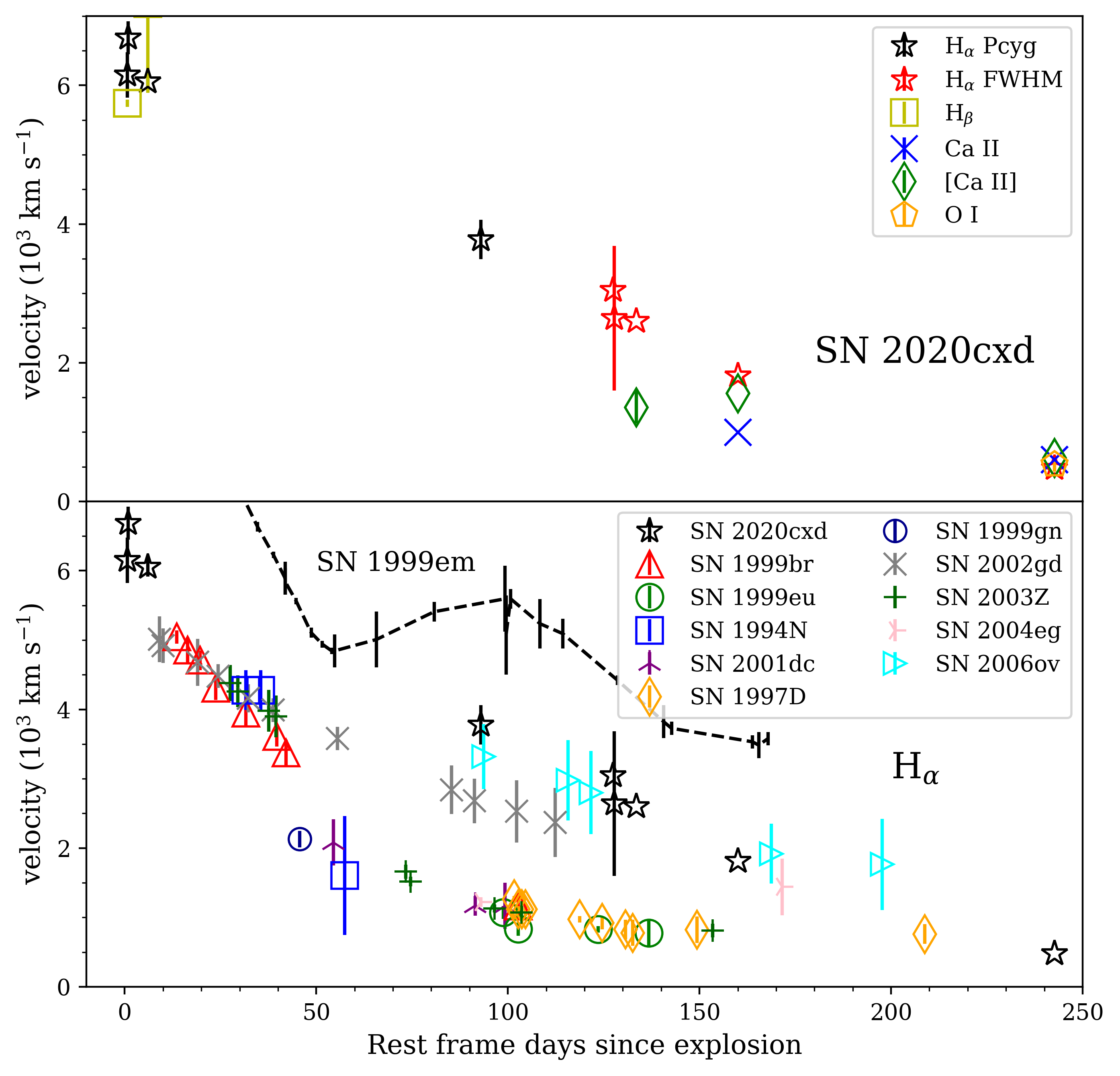}
    \caption{In the upper panel, we show expansion velocities estimated from 
    H$\alpha$ P-Cygni minima (black stars) and FWHM (red stars), 
    H$\beta$ P-Cygni minima (yellow squares),
    [\ion{Ca}{II}]~$\lambda$7291 line FWHM (blue cross),
    FWHM of \ion{Ca}{II} NIR triplet middle line (green diamonds),
    and [\ion{O}{I}]~$\lambda$6300 line FWHM (orange pentagons).
    H$\alpha$ velocities for SN\,2020cxd compared to other LL SNe discussed throughout the paper,
    and comparison with the normal Type II SN 1999em are shown in the lower panel.
    The x-axis shows the phase with respect to explosion.
    Line velocities of the other LL SNe II are taken from \cite{Pastorello04} and \cite{Spiro14}.}
    \label{fig:velocity}
\end{figure*}

The difference between the minimum of the best-fit Gaussian and the line location was translated to an expansion velocity.
In the late nebular phase, 
we instead estimated the velocities from the emission line 
Full Width at Half Maximum (FWHM), which is the width measured at half level between the continuum and the peak of the line, and corrected for the instrumental resolutions obtained from the sky lines.
These velocities are displayed in Fig.~\ref{fig:velocity}, where
we also compare to other LL Type IIP SNe. 
The velocities for the comparison sample are taken from \cite{Pastorello04} and \cite{Spiro14}. 
The time evolution of the velocities measured for H$\alpha$ matches well with those of other LL Type IIP SNe at similar epochs, but also extend to earlier phases. 
The velocities of all these LL SNe, including SN\,2020cxd, are very low, 
clearly lower than for the normal Type IIP SN 1999em. We notice that the H$\alpha$ velocity of SN\,2020cxd became very low around 250 days after the explosion, i.e. we measure an intrinsic FWHM of 478~km~s$^{-1}$, in the last Keck spectrum (Fig.~\ref{fig:spectra_nebula}). This is among the narrowest lines ever measured for a SN IIP.

\begin{figure*}
        \centering
         \includegraphics[width=\textwidth]{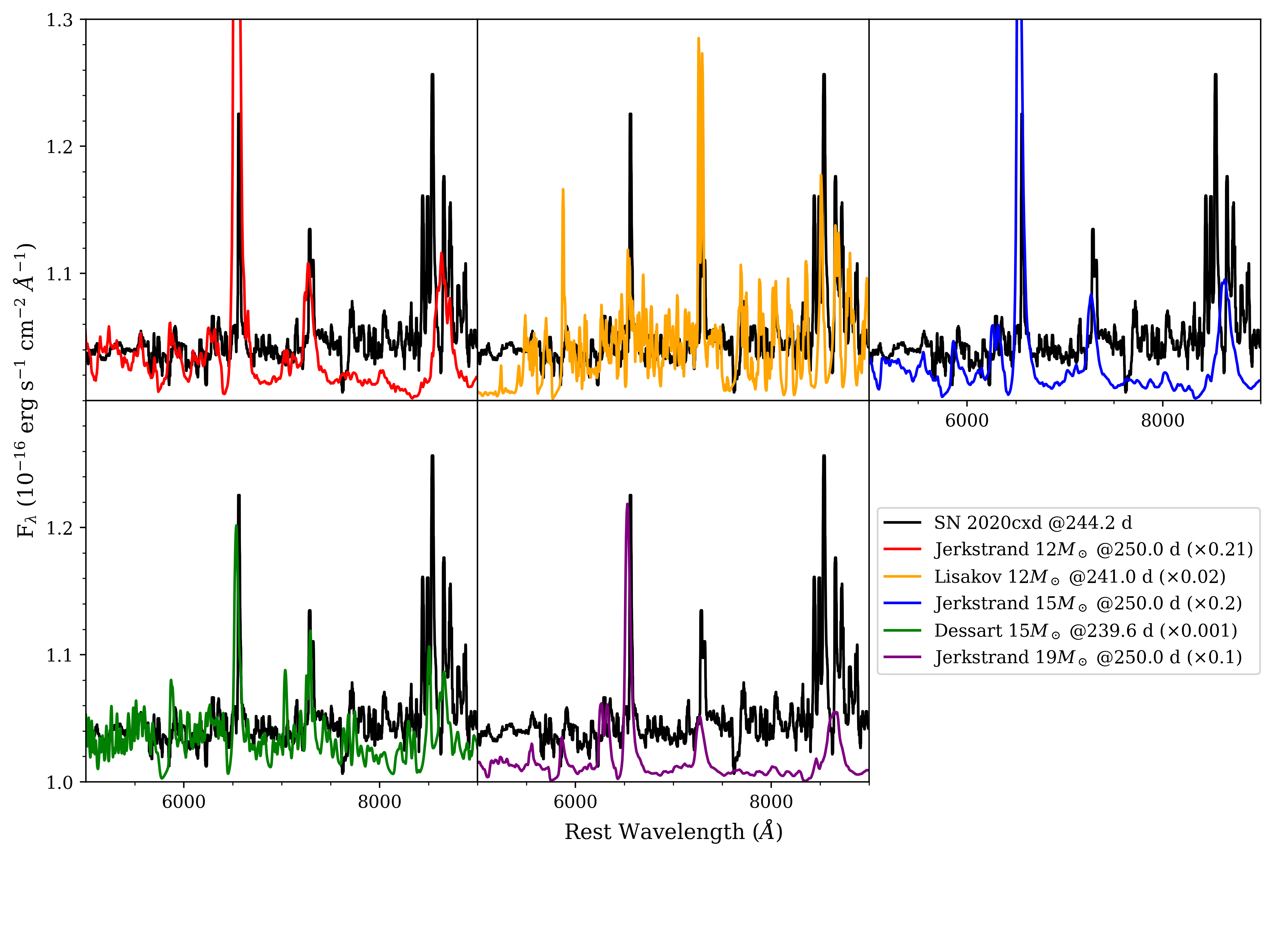}
         \caption{Comparison between the last Keck spectrum of SN\,2020cxd at nebular phases and scaled synthetic spectra. The black curves are the observed spectra, binned with a window size of 10 $\AA$, and corrected for redshift and extinction.
         The models are displayed with colored lines, and originate from \cite{Jerkstrand2012, Jerkstrand2014, Dessart, Lisakov16}. They were obtained via \href{https://wiserep.weizmann.ac.il}{WISeREP}.
         The models are scaled by nickel mass ratio.
         All model spectra are shifted to the SN distance by the inverse square of the distance.
         Their rest frame phases and scaling factors are shown in the legend.}
    \label{fig:spectraid}
\end{figure*}

An important method for diagnosing the progenitor mass is through late-time spectral modeling. At the optically thin phase we are observing deeper into the progenitor structure \citep{Jerkstrand2012}. 
Figure~\ref{fig:spectraid} shows our last Keck spectrum taken during the optically thin phase, compared to modelling.
The Keck spectrum is absolute calibrated against $r$-band interpolated photometry.
In order to estimate the progenitor mass of
SN\,2020cxd, we have collected spectral synthesis models of SNe II (with 12, 15 and 19~\msun~ZAMS masses) from the literature, i.e. from \cite{Lisakov16, Jerkstrand2012, Jerkstrand2014, Dessart}.
The models are selected to have similar phases to the SN observation, and 
are shifted to the distance of SN\,2020cxd, 
as well as scaled by the nickel mass.
As discussed, e.g. in \cite{Jerkstrand2014} and  \cite{Bravo2020}, 
the luminosity of some lines, like [\ion{O}{I}]~$\lambda$6300, 6364,
scales relatively linearly with the ejected nickel mass, thus it is reasonable to scale the model fluxes by nickel mass for the comparison.
To fit the models to the data, we implement a simple Monte Carlo approach that randomly scaled the models while $\chi^2$ values were calculated to quantify the comparisons\footnote{For a specific line, we fit it with a Gaussian and integrated the flux using the trapezoidal rule over the Gaussian range above the continuum.}.
In Fig.~\ref{fig:spectraid}, the five theoretical models with best fit scaling factors are compared to the Keck spectrum of SN\,2020cxd at $\sim 240$~d.
As shown, the 19 \msun~model at this phase is dominated by H$\alpha$ and under-predict the observed fluxes of the other elements.
For the four less massive progenitor models, we see that the models partially reproduce the [\ion{O}{I}]~$\lambda$6300 line.
The Jerkstrand 12 and 15 \msun~models under-predict the [\ion{O}{I}]~$\lambda$6364 by 40\%.
The Dessart 15 \msun~model reproduces the [\ion{Fe}{II}]~$\lambda$7155 of SN\,2020cxd  well, while for the rest of the spectrum, the differences are relatively larger (by $\sim 30-40$\%).
The Jerkstrand models do not reproduce the \ion{Ca}{II} NIR triplet, while the Lisakov 12 \msun~model does a good job reproducing it, better than in the Dessart 15 \msun~model.
The Lisakov model over-predict the
[\ion{Ca}{II}]~$\lambda\lambda$7291, 7323
by a factor of 2, whereas the other models agree better. 
In terms of $\chi^2$ values, the scaled Dessart 15 \msun~model gives the best match, slightly better than the other three models.
Overall, we conclude that the progenitor of SN\,2020cxd had a ZAMS mass
less than 19 \msun. However, to properly distinguish between 12 \msun~and 15 \msun~models, a more detailed comparison is needed. 
We note that for lower-mass progenitors, a smaller line ratio of [\ion{O}{I}]/[\ion{Ca}{II}] is also expected due to the lower O-core mass \citep{Maeda2007, Jerkstrand2012, fang2018}. Using our absolute calibrated Keck spectrum, we measure the fluxes for [\ion{O}{I}]~$\lambda\lambda$6300, 6364 and [\ion{Ca}{II}]~$\lambda\lambda$7291, 7323, and estimate a ratio of 0.82, which favours a lower mass progenitor.

\subsection{Bolometric lightcurve}
\label{sec:bl}

In order to estimate the total radiative output, we attempted to construct a bolometric LC. 
We perform 
Blackbody (BB) fitting every 2 days with the GP interpolated $gri$ data of SN\,2020cxd during its photospheric phase, 
i.e. up to $\sim$ 120 rest frame days.
More details about the diluted BB fitting and the UV/IR contributions to the bolometric LC are discussed in Appendix \ref{sec:bb}.

At nebular phases, the BB fits are not applicable and we 
use the bolometric correction (BC) approach to estimate the bolometric magnitudes of SN\,2020cxd.
In this work, we use the $BC$ of SN~2018hwm to estimate the bolometric LC of SN\,2020cxd. More information on this approach is provided in Appendix \ref{sec:bc}.

\begin{figure*}
\centering
    \includegraphics[width=0.8\textwidth]{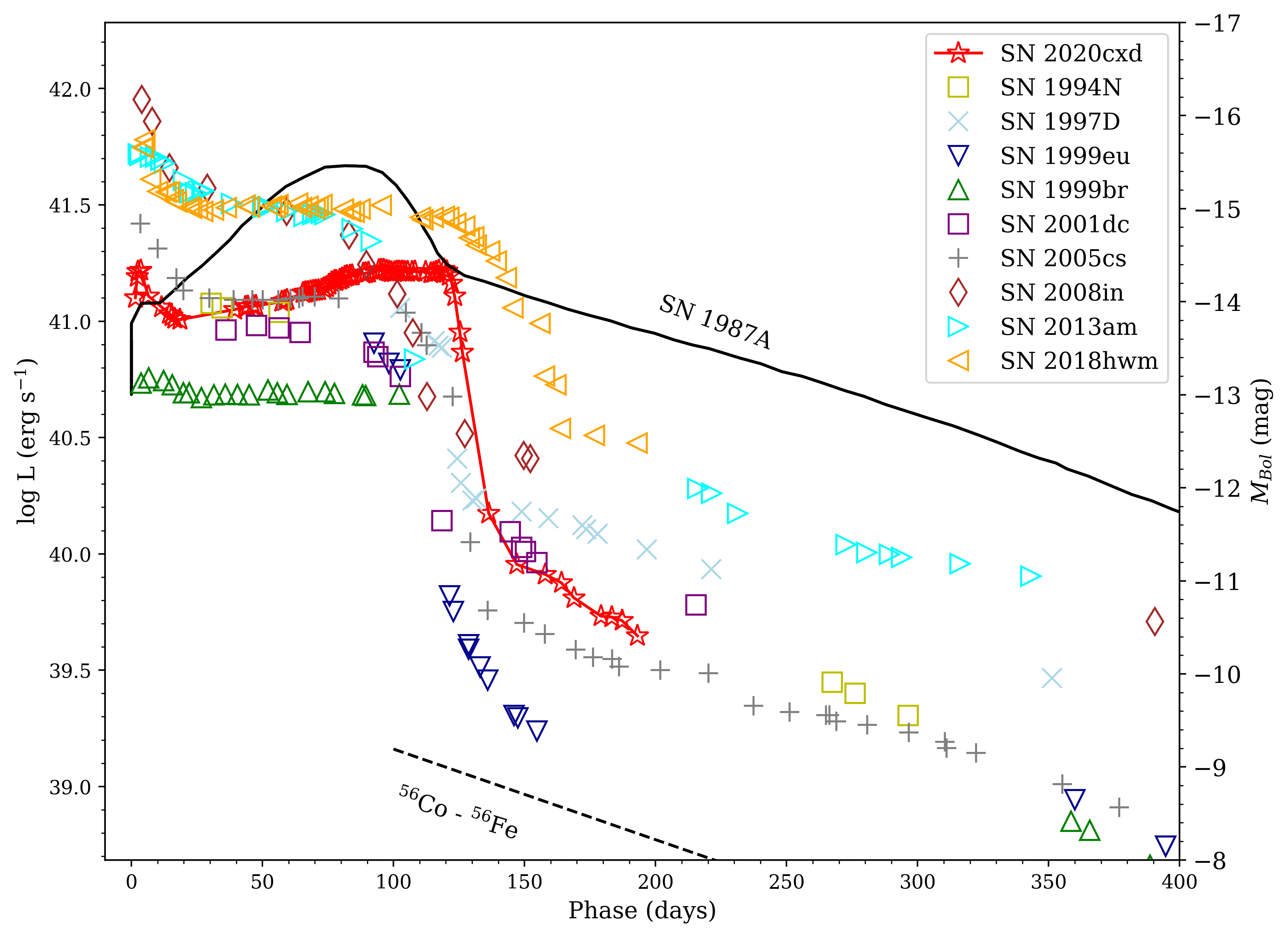}
    \caption{Bolometric luminosities of several LL SNe IIP and the famous SN 1987A. 
    The luminosity of SN\,2020cxd (shown as the red stars) was calculated after accounting for MW extinction and distance. 
    The slope of the luminosity tail matches well with the radioactive decay of $^{56}$Co.
        }
    \label{fig:luminosity}
\end{figure*}

The best fit bolometric luminosity LC of SN\,2020cxd is thus composed of the early part from BB fitting and the tail inferred using the BC, and is shown as red stars in Fig.~\ref{fig:luminosity}, compared to other LL Type IIP SNe discussed throughout the paper. 
As shown, SN\,2020cxd is similar to the other LL SNe II, which are fainter than the famous SN 1987A.
The slope of the LC tail matches well with the Cobalt decay rate.

We can now calculate the bolometric peak luminosity of SN\,2020cxd, which is
L$^{\rm{max}}_{\rm{bol}} = 1.91 \times 10^{41}$ erg s$^{-1}$
at 103 rest frame days
and a total radiated energy over the first 200 rest frame days of
E$_{\rm{rad}} =  1.52 \times 10^{48}$ erg. 

\begin{figure*}
\centering
    \includegraphics[width=0.8\textwidth]{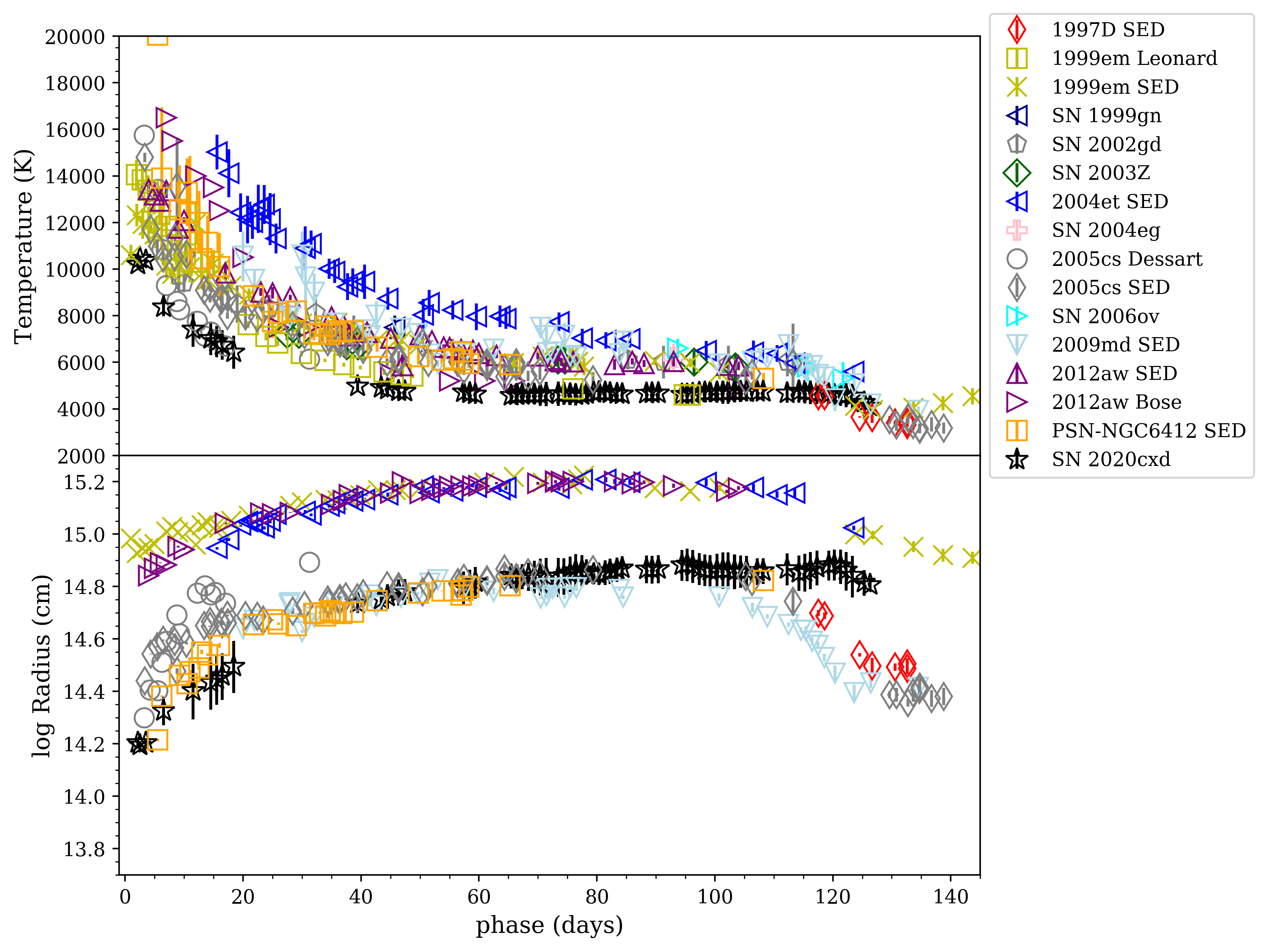}
    \caption{
    Temperature and radius of several LL SNe IIP, including SN\,2020cxd (shown as black stars).
    Upper panel: Evolution of their photospheric temperature estimated by fitting a black-body to the SED in the optically thick phase. 
    Bottom panel: Evolution of their photospheric radius based on the same fits. We have followed \cite{Jager2020} in the methodology, and the included SNe are also from that paper.
    }
    \label{fig:temperature}
\end{figure*}

From the BB approximation we also obtain the temperature and the evolution of the BB radius. The obtained temperatures and radii are compared to those of other LL SNe in Fig.~\ref{fig:temperature}. SN\,2020cxd follows a similar evolution in these parameters as the other LL SNe II in this sample, with both temperature and radius at the lower bound of the distribution. This is sensitive to the amount of host extinction and allowing for additional extinction would increase the temperature.

The ejected $^{56}$Ni mass can be inferred by measuring the luminosity tail, which is powered by the decay of radioactive elements, i.e. $^{56}$Co, synthesized in the explosion.
Using, 
$L  = 1.45 \times 10^{43}$ exp$(-\frac{t}{\tau_{Co}}) (\frac{M_{Ni}}{M_{\sun}}) $ erg s$^{-1}$ from \cite{Nadyozhin} implies that we would require 
$0.002\pm0.001$ \msun~of $^{56}$Ni to account for the tail luminosity (> 140 days).
We note that the uncertainty in distance modulus affects this estimate with a systematic uncertainty of $\lesssim 30$\%, but an additional host extinction of $E(B-V)_{\rm{host}} = 0.25$~mag would increase this by a factor two.

In order to estimate progenitor and explosion parameters from the photometry, we make use of a Monte Carlo code that was recently presented by \citet{Jager2020}, which fits semi-analytic models to quasi-bolometric LCs (see Appendix \ref{sec:mc} for more details).
We employ the MCMC code to fit the semi-analytic models to SN\,2020cxd, the acceptable fits are displayed as light solid black lines in Fig.~\ref{fig:lumcompare}.
After marginalization, we have estimates with confidence intervals (1$\sigma$) for each of these parameters, i.e.
SN\,2020cxd has $R_0=1.3_{-0.1}^{+0.4}\times10^{13}$ cm 
($\sim187~R_\odot$), 
$M_{ej}=9.5_{-1.0}^{+1.3}$ \msun,
$E_{kin}=4.3_{-0.9}^{+0.9}\times10^{50}$ erg, 
$v_{exp}=2747_{-235}^{+714}$ km s$^{-1}$,
$E_{th}=1.5_{-0.5}^{+0.3} \times10^{50}$ erg,
for its progenitor radius, ejecta mass, kinetic energy, expansion velocity, and thermal energy respectively.
The nickel mass was simultaneously estimated as 
$0.003\pm0.002$ \msun.

\cite{Reguitti2020} estimated the ejected mass of SN~2018hwm to 
8~\Msun. By assuming the mass of the compact remnant and a typical mass loss during the pre-SN stage, they estimated that the initial progenitor mass 
(zero age main sequence, ZAMS mass)
of SN 2018hwm was in the range of $9.4-10.9$ \Msun.
For SN\,2020cxd, the ejected mass is estimated to be $\sim9.5$ \Msun.
With the same assumptions, its ZAMS mass was then between 10 and 14 \Msun.

The initial bolometric decline (first 20 days) somewhat mimics those suggested to be powered by CSM in some hydrodynamical models \citep{Morozova2017}, but no evidence was found for flash spectroscopy features \citep[e.g.,][]{Bruch_2020}
in our very early spectra (< 3 days past explosion). Overall, there is no evidence for CSM interaction from narrow spectral lines, UV brightness or from the shape of the LC. These consideration could disfavour a SAGB star as a potential progenitor star, since these are thought to have high mass-loss rates  \citep[compared to those of RSG stars of similar initial mass,][]{Hiramatsu2020}.

There has been extensive studies on the duration of the LC plateau for Type II SNe in the literature, exploring for example the effect of radioactivity on this phase. Although well established that the luminosity in this period derives from the diffusion as the photosphere recedes, it has also been suggested that the duration of the plateau is affected by radioactive input \citep{Popov, Kasen_2009, Nakar2016}.
The most important parameter for the plateau duration in the investigation 
of \cite{Pumo16} 
was the amount of stripping of the ejecta, where extensive stripping led to shorter plateau lengths. In this context, the very long duration in SN\,2020cxd might suggest that the ejecta must have remained relatively intact. 
\cite{Kasen_2009} derived scaling relations for the properties of
SNe IIP, and conclude 
that the plateau duration is correlated with the explosion energy and the progenitor mass 
(where nickel in the ejecta tends to extend the plateau).
However, given the low plateau luminosities of SN 2020cxd and SN 2018hwm these scaling relations would suggest a too low initial mass; this 
discrepancy was explained for SN 2018hwn by the extremely low explosion energy \citep{Reguitti2020}\footnote{Note, however, that their original value for the explosion energy of SN 2018hwm is too low due to a numerical error (private communication, see also their errata from April 2021); their errata suggests 0.075 foe, our MCMC estimate is instead 0.23 foe for SN 2018hwn.}.

\begin{figure*}
\centering
    \includegraphics[width=0.8\textwidth]{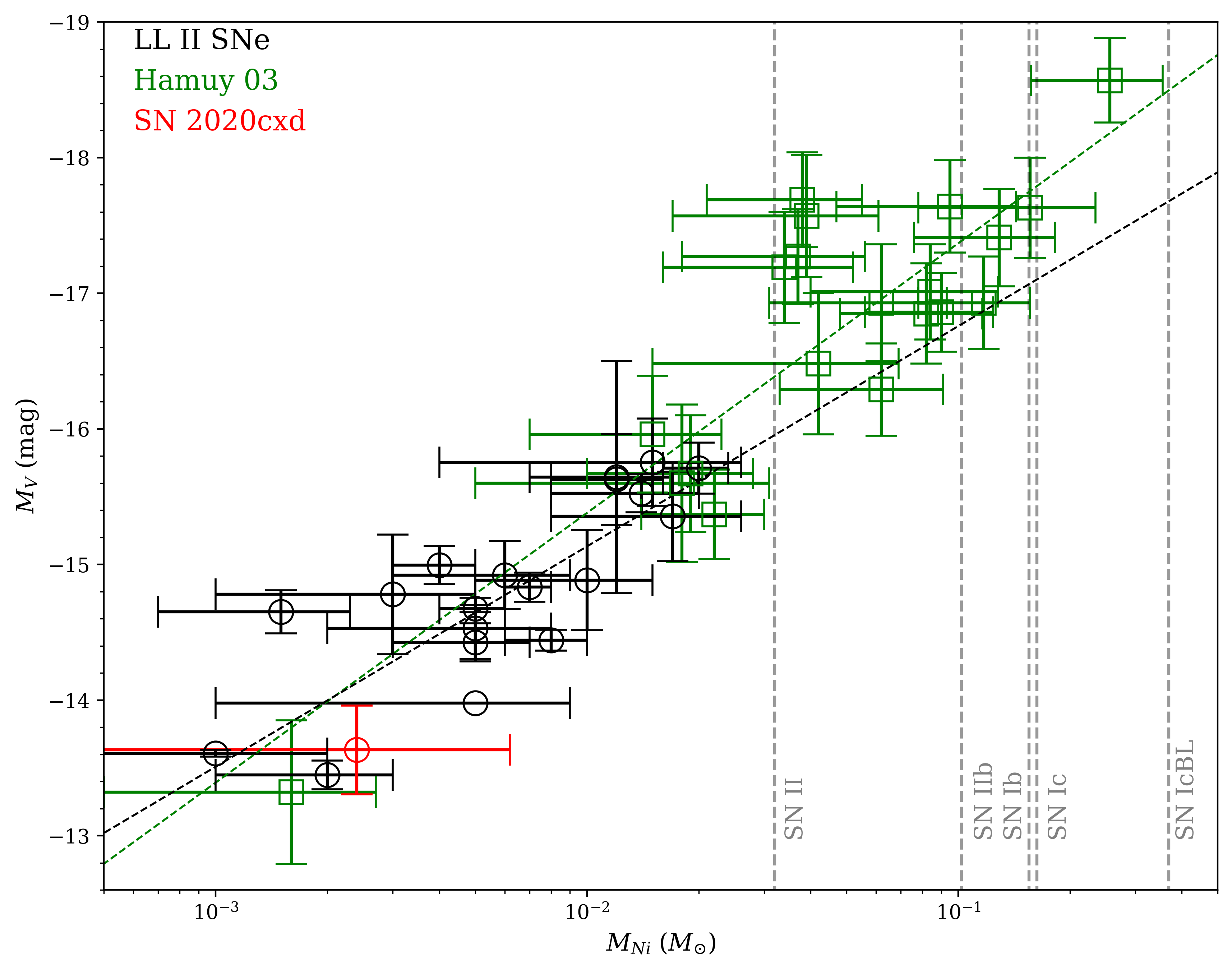}
    \caption{ 
    Absolute $V$-band plateau magnitudes (computed at day 50) versus
    ejected nickel mass.
    Black circles represent the LL SN IIP sample (from Table \ref{tab:llsn}), green squares are the SNe IIP from \cite{Hamuy2003}.
    SN\,2020cxd is shown as a red data point.
    The grey dashed vertical lines represent the average nickel mass for different types of CC SNe from \cite{Anderson2019}. 
    We perform linear fits on $M_V$ and log$M_{Ni}$ for the Hamuy 2003 sample and for the LL SN IIP sample. The best fits are shown as the green and black dashed lines, and give consistent relations.}
    \label{fig:ni}
\end{figure*}

In Fig.~\ref{fig:ni} we compare the estimated radioactive $^{56}$Ni mass of SN\,2020cxd to those estimated for LL SNe II in the literature. As shown, the $^{56}$Ni mass of SN\,2020cxd is significantly lower than for normal CC SNe, and in the range of the LL SN class.
We also examine the correlation between the $V$ band ($g$ band for our case) absolute magnitude and the ejected nickel mass.
As shown, the linear relation of \cite{Hamuy2003} can be confirmed also for the LL Type II SNe, extending the relation to the lower nickel mass region.
There is evidence both from the photospheric velocities (Fig.~\ref{fig:velocity}), the very narrow nebular emission lines (Fig.~\ref{fig:spectra_nebula}) and the LC modeling that the explosion has very low energy. A priori, there need not be any correlation between the explosion energy, the luminosity at the plateau (Fig.~\ref{fig:ni}) and the ejected mass of radioactive nickel, but this SN adds to the evidence that such correlations exist.
Explosion models of neutrino-driven Fe CC indeed predict that in this ZAMS mass regime, 
where the iron core mass scales with ZAMS mass, lower explosion energies and thus ejecta velocities are expected from lower ZAMS mass stars, which also eject less mass \citep{2021arXiv210201118B}.

\subsection{Host galaxy properties}
\label{sec:host_analysis}

The host galaxy of SN 2020cxd, NGC 6395, is classified as a Scd galaxy in NED, and reported to have a stellar mass $\log M_* = 9.29 \pm 0.10$~M$_{\odot}$ by \citet{Leroy2019a}. This is close to the center of the distribution of galaxy masses for SNe II hosts measured by \citet{Schulze_2020} for the 
Palomar Transient Factory (PTF)
sample (median $\log M_* = 9.65$~M$_{\odot}$). Also many of the other LL SNe have been discovered in large spirals \citep{Spiro14}. 

No metallicity measurement for NGC 6495 is reported in the literature. As described in Sect.~\ref{sec:host}, we therefore extracted the spectrum of a nearby H~II region from the day 242 Keck spectrum in order to obtain a metallicity estimate. We measure the fluxes of a range of emission lines (Table~\ref{tab:galflux}) and
use the flux ratios to calculate several common strong-line metallicity indicators using the {\tt pyMCZ} code \citep{Bianco2016}. The Balmer decrement indicates moderate extinction, $E(B-V) = 0.36 \pm 0.02$ mag towards this H II region. We measure a slightly sub-solar metallicity in most indicators, for example we can adopt the N2 scale of the \cite{Pettini_2004} calibration using the flux ratio between [\ion{N}{II}] $\lambda$6583 and H$\alpha$. We found that $12+{\rm log (O/H)}=8.50\pm0.15$.
We also employed another metallicity diagnostic of \cite{Dopita_2016} using [\ion{N}{II}], H$\alpha$ and [\ion{S}{II}] lines, in which $12+{\rm log(O/H)}=8.40\pm0.10$.   
Assuming a solar abundance of 8.69 \citep{Asplund_2009}, the oxygen abundance of the host galaxy is  $0.63^{+0.26}_{-0.12}$\,$Z_{\odot}$. Comparing to the stellar mass estimate of $\sim10^{9.29}\,M_{\odot}$ from  \citet{Leroy2019a}, our metallicity is consistent with the galaxy mass-metallicity relation (e.g. \citealt{Andrews_2013}).

\section{Summary and Conclusions}
\label{sec:summary}

We have presented SN\,2020cxd, a LL SN IIP discovered by the ZTF which has one of the longest plateaus, the slowest ejecta and the coolest photospheres among such objects. The LC is well-sampled with a well determined explosion epoch and shows an initial sharp rise, and then a negative slope ($s_2$) during the plateau phase which peaks after about 100 days. The long-duration (118 days, OPTd) plateau length ends with a very sharp drop to a low-luminosity radioactive tail, implying
0.003 \msun~of
$^{56}$Ni being ejected in the explosion.
We find no evidence for a precursor at the site of the explosion, which is measured to have slightly sub-solar metallicity. 


We compare the nebular spectra to spectral synthesis models to constrain the progenitor mass through the [\ion{O}{I}]~$\lambda$6300, 6364 lines, and find relatively good agreements with progenitors of 12 and 15 \msun, excluding more massive progenitor scenarios. 
This argues against a high mass ($25-40$ \msun) RSG (as a Fe CC SN) with large amounts of material from the Ni-rich region falling back onto the newly formed degenerate core \citep{Zampieri03}.

We constructed the bolometric LC of SN\,2020cxd, and estimated a total radiated energy of
E$_{\rm{rad}} =  9.71 \times 10^{49}$ erg
over the first 200 days. 
The semi-analytic modelling suggests that SN\,2020cxd originates from a progenitor which ejected 
about 9.5 \msun~of ejecta.
Although we can not conclusively determine the explosion mechanism (Electron capture or Fe CC), or the nature of the progenitor, all evidence are consistent with a Fe CC of a progenitor with a
ZAMS mass on the order of  12 \msun.


This presentation provides an observational account of SN 2020cxd, which deserved a detailed investigation given its unique position in the larger systematic BTS survey. We have argued that this SN is most likely a result from a CC in a low ZAMS mass star. This is in accordance with the notion that the low-luminosity tail of the SN luminosity function is underexplored, and that understanding the SN progenitor IMF require further studies of large 
well-controlled samples in tandem with detailed studies of individual examples to connect them to the stellar progenitors.

\begin{acknowledgements}

Based on observations obtained with the Samuel Oschin Telescope 48-inch and the 60-inch Telescope at the Palomar Observatory as part of the Zwicky Transient Facility project. ZTF is supported by the National Science Foundation under Grant No. AST-1440341 and a collaboration including Caltech, IPAC, the Weizmann Institute for Science, the Oskar Klein Center at Stockholm University, the University of Maryland, the University of Washington, Deutsches Elektronen-Synchrotron and Humboldt University, Los Alamos National Laboratories, the TANGO Consortium of Taiwan, the University of Wisconsin at Milwaukee, and Lawrence Berkeley National Laboratories. Operations are conducted by COO, IPAC, and UW.  The SED Machine is based upon work supported by the National Science Foundation under Grant No. 1106171. This work was supported by the GROWTH \citep{growth} project funded by the National Science Foundation under Grant No 1545949. Partially based on observations made with the Nordic Optical Telescope, operated at the Observatorio del Roque de los Muchachos, La Palma, Spain, of the Instituto de Astrofisica de Canarias. Some of the data presented here were obtained with ALFOSC, which is provided by the Instituto de Astrofisica de Andalucia (IAA) under a joint agreement with the University of Copenhagen and NOTSA. 
This research has made use of the SIMBAD database,
operated at CDS, Strasbourg, France, and of the NASA/IPAC Extragalactic Database (NED),
which is operated by the Jet Propulsion Laboratory, California Institute of Technology,
under contract with the National Aeronautics and Space Administration.
MMK acknowledges generous support from the David and Lucille Packard Foundation.
SY acknowledge support from the G.R.E.A.T research environment, funded by {\em Vetenskapsr\aa det}, the Swedish Research Council, project number 2016-06012.
MR has received funding from the European Research Council (ERC) under the European Union's Horizon 2020 research and innovation programme (grant agreement no. 759194 - USNAC). We thank Peter Nugent for comments on the manuscript. We thank the referee for a thorough read, and a reminder about the negative plateau slope seen in many LL SNe II.

\end{acknowledgements}

\bibliography{ref}

\onecolumn

\begin{deluxetable}{lccccccccc}
\tablewidth{0pt}
\tabletypesize{\scriptsize}
\tablecaption{Properties of the sample of LL SNe IIP. The sample was composed by \cite{Spiro14} and \cite{Bravo2020}, and was extended with 3 recent objects, i.e. SN-NGC 6412, SN 2018hwm and SN\,2020cxd. 
$t_0$ is the estimated explosion time, which was used to shift the 
abscissa in Fig.~\ref{fig:lc+iptf}. 
References: (1) \cite{Pastorello04}; (2) \cite{Turatto_1998}; (3) \cite{Benetti_2001}; (4) \cite{Spiro14}; (5) \cite{Anderson_2014}; (6) \cite{Galbany2016}; (7) \cite{Gutierrez2017}; 
(8) \cite{Pastorello_2006}; (9) \cite{Pastorello_2009}; (10) \cite{Mattila08}; (11) \cite{VanDyk}; (12) \cite{roy2011}; (13) \cite{takats2014}; (14) \cite{Fraser2011}; (15) \cite{Gal_Yam_2011}; (16) \cite{Zhang2014}; (17) \cite{Jager2020}; (18) \cite{Nakaoka_2018}; (19) \cite{Bravo2020}; (20) \cite{Reguitti2020}; (21) this paper.
\label{tab:llsn}}
\tablehead{
\colhead{SN name} &
\colhead{$t_0$} & 
\colhead{z} & 
\colhead{$\mu$} &
\colhead{$A_v$(MW)} &
\colhead{$A_v$(Host)} &
\colhead{Host} &
\colhead{$M_{Ni}$} &
\colhead{Ref}\\
\colhead{} &
\colhead{(MJD)}  & 
\colhead{}  & 
\colhead{(mag)} &
\colhead{(mag)} &
\colhead{(mag)} &
\colhead{} &
\colhead{(\msun)} &
\colhead{} 
}
\startdata
SN 1994N & 49451 &  0.0098 & $33.09\pm0.31$ & 0.108 & 0.000 & UGC 5695 & $0.005\pm0.001$ & 1,4 \\
SN 1997D & 50361 & 0.004059 & $30.74\pm0.92$  & 0.058 & $\lesssim0.060$ & NGC 1536 & $0.005\pm0.004$ & 2,3,4 \\
SN 1999br & 51278 & 0.00323 & $30.97\pm0.83$  & 0.065 & 0.000 & NGC 4900 & $0.002\pm0.001$ & 1 \\
SN 1999eu & 51394 & 0.0042 & $30.85\pm0.87$  & 0.073 & 0.000 & NGC 1097 & $0.001\pm0.001$ & 1,4 \\
SN 2001dc & 52047 & 0.0071 & $32.64\pm0.38$  & 1.654 & 0.046 & NGC 5777 & $0.005\pm0.002$ & 1 \\
SN 2002gd & 52552 & 0.00892 & $32.87\pm0.35$  & 0.184 & 0.000 & NGC 7537 & $\lesssim0.003$ & 4 \\
SN 2002gw & 52568 & 0.01028 & $32.98\pm0.23$ & 0.051 & 0.000 & NGC 922 & $0.012\pm0.004$ & 5,6,7 \\
SN 2003B & 52645 & 0.00424 & $31.11\pm0.28$ & 0.072 & 0.180 & NGC 1097 & $0.017\pm0.009$ & 5,6,7 \\
SN 2003fb & 52797 & 0.01754 & $34.43 \pm0.12$ & 0.482 & 0.000 & UGC 11522 & $\gtrsim0.017$ & 5,6,7 \\
SN 2003Z & 52665 & 0.0043 & $31.70\pm0.60$  & 0.106 & 0.000 & NGC 2742 & $0.005\pm0.003$ & 4 \\
SN 2004eg & 53170 & 0.008051 & $32.64\pm0.38$  & 1.237 & 0.000 & UGC 3053 & $0.007\pm0.003$ & 4 \\
SN 2004fx & 53281 & 0.00892 & $32.82\pm0.24$ & 0.274 & 0.000 & MCG -02-14-003 & $0.014\pm0.006$ & 5,7 \\
SN 2005cs & 53549 & 0.002 & $29.46\pm0.60$  & 0.095 & 0.171 & M 51 & $0.006\pm0.003$ & 8,9 \\
SN 2006ov & 53974 & 0.0052 & $30.5\pm0.95$  & 0.061 & 0.000 & NGC 4303 & $0.002\pm0.002$ & 4 \\
SN 2008bk & 54550 & 0.000767 & $27.68\pm0.13$ & 0.065 & 0.000 & NGC 7793 & $0.007\pm0.001$ & 10,11 \\
SN 2008in & 54825 & 0.005224 & $30.60\pm0.20$  & 0.305 & 0.080 & NGC 4303 & $0.012\pm0.005$ & 12 \\
SN 2009N & 54848 & 0.003456 & $31.67\pm0.11$  & 0.350 & 0.100 & NGC 4487 & $0.020\pm0.004$ & 13 \\
SN 2009md & 55162 & 0.00427 & $31.64\pm0.21$  & 0.310 & 0.000 & NGC 3389 & $0.004\pm0.001$ & 14 \\
SN 2010id & 55452 & 0.01648 & $32.86\pm0.50$  & 0.162 & 0.167 & NGC 7483 & - & 15 \\
SN 2013am & 56345 & 0.002692 & $30.54\pm0.40$ & 0.066 & 1.705 & NGC 3623 & $0.015\pm0.011$ & 16 \\
SN-NGC 6412 & 57210 & 0.00438 & $22.18\pm1.56$ & 0.115 & 0.000 & NGC 6412 & $0.002\pm0.001$ & 17 \\
SN 2016bkv & 57477 & 0.002 & $30.79\pm0.05$ & 0.045 & $\lesssim0.016$ & NGC 3184 & $0.0216\pm0.0014$ & 18 \\
SN 2016aqf & 57444 & 0.004016 & $30.16\pm0.27$ & 0.146 & $\lesssim0.096$ & NGC 2101 & $0.008\pm0.002$ & 19 \\
SN 2018hwm & 58425 & 0.00895 & $33.58\pm0.19$ & 0.071 & 0.000 & IC 2327 & $0.003\pm0.002$ & 20 \\
\textbf{SN 2020cxd} & 58897 & 0.0039 & $31.70\pm0.30$ & 0.115 &0.000 & NGC 6395 & $0.002\pm0.001$ & 21 \\
\enddata
\end{deluxetable}

\begin{deluxetable}{lccccccc}
\tablewidth{0pt}
\tabletypesize{\scriptsize}
\tablecaption{Summary of ground-based photometry for SN\,2020cxd \label{tab:ground}}
\tablehead{
\colhead{Observation Date} & 
\colhead{Rest Frame Phase} & 
\colhead{Telescope} & 
\colhead{u} & 
\colhead{g} & 
\colhead{r} &
\colhead{i} &
\colhead{z} \\
\colhead{(JD)} &
\colhead{(day)}  & 
\colhead{}  & 
\colhead{(mag)} &
\colhead{(mag)} &
\colhead{(mag)} &
\colhead{(mag)} &
\colhead{(mag)}
}
\startdata
2458896.01 & -1.52 & P48 & - & $\gtrsim$ 20.39 & $\gtrsim$ 20.26 & - & - \\
2458899.04 & 1.51 & P48 & - & 17.63 (0.04) & 17.69 (0.05) & - & - \\
2458899.72 & 2.24 & LT & 17.69 (0.01) & 17.63 (0.01) & 17.64 (0.01) & 17.78 (0.01) & 17.90 (0.02) \\
2458900.00 & 2.46 & P48 & - & 17.61 (0.05) & 17.65 (0.05) & - & - \\
2458901.01 & 3.47 & P48 & - & 17.66 (0.07) & 17.62 (0.05) & - & - \\
2458904.02 & 6.47 & P48 & - & 17.75 (0.05) & 17.59 (0.05) & - & - \\
2458909.02 & 11.45 & P48 & - & 17.86 (0.06) & 17.55 (0.05) & - & - \\
2458912.02 & 14.43 & P48 & - & 17.97 (0.07) & 17.58 (0.06) & - & - \\
2458913.01 & 15.42 & P48 & - & 17.98 (0.07) & 17.59 (0.06) & - & - \\
2458914.01 & 16.41 & P48 & - & 18.04 (0.07) & 17.62 (0.07) & - & - \\
2458915.05 & 17.45 & P48 & - & - & 17.63 (0.05) & - & - \\
2458915.99 & 18.39 & P48 & - & 18.08 (0.08) & 17.34 (0.19) & - & - \\
2458937.00 & 39.32 & P48 & - & 18.33 (0.07) & 17.54 (0.04) & - & - \\
2458940.99 & 43.29 & P48 & - & 18.29 (0.07) & - & - & - \\
2458942.00 & 44.29 & P48 & - & 18.39 (0.08) & - & - & - \\
2458943.95 & 46.24 & P48 & - & 18.40 (0.13) & 17.50 (0.05) & - & - \\
2458945.00 & 47.29 & P48 & - & 18.39 (0.09) & 17.51 (0.05) & - & - \\
2458954.95 & 57.20 & P48 & - & 18.38 (0.08) & 17.44 (0.04) & - & - \\
2458955.93 & 58.17 & P48 & - & 18.39 (0.07) & 17.46 (0.05) & 17.28 (0.06) & - \\
2458956.86 & 59.10 & P48 & - & 18.51 (0.19) & 17.44 (0.05) & - & - \\
2458961.95 & 64.17 & P48 & - & - & 17.40 (0.05) & - & - \\
2458962.98 & 65.20 & P48 & - & 18.40 (0.10) & 17.40 (0.04) & - & - \\
2458963.94 & 66.15 & P48 & - & 18.32 (0.07) & - & - & - \\
2458964.90 & 67.11 & P48 & - & 18.34 (0.07) & 17.38 (0.04) & - & - \\
2458965.94 & 68.15 & P48 & - & 18.33 (0.06) & 17.37 (0.04) & 17.16 (0.04) & - \\
2458966.95 & 69.15 & P48 & - & 18.30 (0.07) & 17.34 (0.04) & - & - \\
2458967.92 & 70.12 & P48 & - & 18.35 (0.08) & 17.31 (0.05) & - & - \\
2458968.92 & 71.12 & P48 & - & 18.33 (0.06) & 17.33 (0.04) & - & - \\
2458969.91 & 72.10 & P48 & - & - & 17.27 (0.09) & 17.31 (0.26) & - \\
2458970.93 & 73.12 & P48 & - & 18.30 (0.06) & 17.29 (0.04) & - & - \\
2458971.93 & 74.11 & P48 & - & 18.27 (0.07) & 17.30 (0.05) & - & - \\
2458972.95 & 75.12 & P48 & - & 18.22 (0.06) & 17.27 (0.04) & - & - \\
2458973.90 & 76.07 & P48 & - & 18.30 (0.08) & 17.25 (0.05) & 17.08 (0.05) & - \\
2458974.90 & 77.07 & P48 & - & 18.24 (0.09) & 17.21 (0.04) & - & - \\
2458975.85 & 78.01 & P48 & - & 18.17 (0.08) & 17.28 (0.04) & - & - \\
2458976.91 & 79.07 & P48 & - & 18.17 (0.09) & 17.26 (0.04) & - & - \\
2458977.88 & 80.04 & P48 & - & 18.06 (0.08) & 17.24 (0.05) & 17.01 (0.04) & - \\
2458978.87 & 81.02 & P48 & - & 18.12 (0.10) & 17.19 (0.04) & - & - \\
2458979.80 & 81.95 & P48 & - & 18.13 (0.07) & 17.19 (0.05) & - & - \\
2458980.94 & 83.09 & P48 & - & 18.08 (0.10) & 17.21 (0.04) & - & - \\
2458981.87 & 84.01 & P48 & - & 18.18 (0.10) & 17.20 (0.04) & 16.98 (0.05) & - \\
2458985.92 & 88.05 & P48 & - & 18.10 (0.07) & 17.16 (0.04) & - & - \\
2458986.91 & 89.03 & P48 & - & 18.09 (0.05) & 17.17 (0.05) & 16.97 (0.04) & - \\
2458987.89 & 90.01 & P48 & - & 18.09 (0.06) & 17.15 (0.04) & - & - \\
2458991.85 & 93.95 & P48 & - & 18.07 (0.06) & 17.13 (0.04) & - & - \\
2458992.84 & 94.95 & P48 & - & 18.14 (0.06) & 17.11 (0.04) & 16.94 (0.04) & - \\
2458993.76 & 95.86 & P48 & - & 18.08 (0.07) & - & - & - \\
2458994.90 & 96.99 & P48 & - & 18.04 (0.05) & - & - & - \\
2458995.80 & 97.89 & P48 & - & 18.09 (0.06) & 17.12 (0.05) & - & - \\
2458996.78 & 98.86 & P48 & - & 18.05 (0.06) & 17.14 (0.05) & 16.97 (0.05) & - \\
2458997.84 & 99.92 & P48 & - & 18.08 (0.05) & 17.13 (0.04) & - & - \\
2458998.85 & 100.93 & P48 & - & 18.07 (0.05) & 17.11 (0.03) & - & - \\
2458999.79 & 101.87 & P48 & - & 18.10 (0.07) & 17.10 (0.04) & - & - \\
2459000.86 & 102.93 & P48 & - & 18.05 (0.10) & 17.12 (0.05) & - & - \\
2459001.88 & 103.95 & P48 & - & 18.06 (0.07) & - & - & - \\
2459002.91 & 104.97 & P48 & - & 18.11 (0.09) & - & - & - \\
2459004.79 & 106.84 & P48 & - & 18.02 (0.09) & 17.14 (0.04) & 16.95 (0.04) & - \\
2459005.75 & 107.81 & P48 & - & 18.04 (0.07) & 17.14 (0.04) & - & - \\
2459008.87 & 110.91 & P48 & - & - & 17.13 (0.05) & 16.95 (0.04) & - \\
2459009.83 & 111.86 & P48 & - & 18.11 (0.08) & 17.13 (0.05) & - & - \\
2459011.82 & 113.85 & P48 & - & 18.04 (0.07) & 17.13 (0.04) & - & - \\
2459012.86 & 114.88 & P48 & - & 18.07 (0.05) & 17.11 (0.04) & 17.00 (0.05) & - \\
2459013.78 & 115.80 & P48 & - & 18.09 (0.06) & 17.11 (0.03) & - & - \\
2459014.80 & 116.82 & P48 & - & 18.09 (0.06) & 17.11 (0.04) & - & - \\
2459015.82 & 117.84 & P48 & - & - & 17.11 (0.04) & - & - \\
2459016.84 & 118.85 & P48 & - & 18.04 (0.08) & - & - & - \\
2459017.80 & 119.80 & P48 & - & 18.12 (0.05) & 17.14 (0.03) & - & - \\
2459018.83 & 120.83 & P48 & - & 18.22 (0.07) & - & 17.00 (0.04) & - \\
2459019.83 & 121.82 & P48 & - & 18.27 (0.07) & 17.27 (0.04) & - & - \\
2459020.82 & 122.81 & P48 & - & 18.36 (0.08) & 17.34 (0.04) & - & - \\
2459022.80 & 124.79 & P48 & - & 18.99 (0.10) & 17.87 (0.05) & 17.59 (0.04) & - \\
2459023.82 & 125.80 & P48 & - & 19.44 (0.14) & 18.19 (0.06) & - & - \\
2459025.68 & 127.66 & P48 & - & - & 18.91 (0.11) & - & - \\
2459026.79 & 128.76 & P60 & - & - & 19.27 (0.24) & - & - \\
2459032.74 & 134.68 & P60 & - & - & 20.45 (0.42) & - & - \\
2459033.88 & 135.82 & P60 & - & - & 20.44 (0.33) & 19.79 (0.34) & - \\
2459044.59 & 146.49 & LT & - & $\gtrsim$ 22.25 & 21.09 (0.12) & 20.29 (0.04) & - \\
2459055.45 & 157.31 & LT & - & $\gtrsim$ 22.29 & 21.34 (0.16) & 20.49 (0.07) & - \\
2459060.91 & 162.75 & P60 & $\gtrsim$ 19.40 & $\gtrsim$ 19.85 & $\gtrsim$ 19.17 & $\gtrsim$ 19.40 & - \\
2459061.58 & 163.42 & LT & - & - & 21.40 (0.15) & 20.57 (0.10) & 20.10 (0.12) \\
2459066.40 & 168.22 & LT & - & - & - & 20.60 (0.07) & - \\
2459076.72 & 178.50 & P60 & - & - & - & 20.70 (0.23) & - \\
2459080.77 & 182.53 & P60 & - & - & - & 20.72 (0.23) & - \\
2459084.73 & 186.48 & P60 & - & - & - & 20.72 (0.28) & - \\
2459090.68 & 192.40 & P60 & - & - & - & 20.96 (0.24) & - \\
2459092.36 & 193.40 & LT  & - & $\gtrsim$ 21.80 & $\gtrsim$ 20.92 & $\gtrsim$ 20.98 & - \\
2459121.39 & 222.99 & NOT & - & - & 21.80 (0.12) & 21.06 (0.09) & - \\
2459130.34 & 231.92 & LT & - & - & $\gtrsim$ 21.74 & - & - \\
2459131.37 & 232.94 & NOT & - & - & 22.06 (0.11) & 21.24 (0.08) & - \\
\enddata
\end{deluxetable}

\begin{deluxetable}{lcccccccc}
\tablewidth{0pt}
\tabletypesize{\scriptsize}
\tablecaption{Summary of Swift Observations for SN\,2020cxd. Fluxes with SNR less than 3 sigma are shown as upper limits.
\label{tab:swift}}
\tablehead{
\colhead{Observation Date} & 
\colhead{Rest Frame Phase} & 
\colhead{V} & 
\colhead{B} & 
\colhead{U} &
\colhead{UVW1} &
\colhead{UVM2} &
\colhead{UVW2} \\
\colhead{(JD)} &
\colhead{(days)}  & 
\colhead{(mag)} &
\colhead{(mag)} &
\colhead{(mag)} &
\colhead{(mag)} &
\colhead{(mag)} &
\colhead{(mag)} 
}
\startdata
2458899.87 & 2.33 & > 18.04 & 17.53 (0.13) & 17.81 (0.12) & 17.90 (0.11) & 18.54 (0.14) & 18.53 (0.13) \\ 
2458902.39 & 4.84 & 17.87 (0.35) & 17.82 (0.17) & 18.03 (0.15) & 18.44 (0.14) & 19.23 (0.18) & 19.48 (0.23) \\
2458907.62 & 10.05 & 17.62 (0.31) & 17.89 (0.19) & 18.72 (0.25) & > 19.82 & > 20.28 & > 20.21  \\
2458910.15 & 12.57 & > 17.89 & 17.77 (0.18) & > 19.31 & > 19.85 & > 20.38 & > 20.20 \\
2458911.94 & 14.35 & 17.67 (0.32) & 17.79 (0.18) & 19.21 (0.35) & > 19.80  & > 20.43  & > 20.14\\
2459003.04 & 105.10 & > 17.89 & 18.34 (0.26) & > 19.43 & - & - & -  \\
\enddata
\end{deluxetable}

\begin{deluxetable}{lcccc}
\tablewidth{0pt}
\tabletypesize{\scriptsize}
\tablecaption{Summary of Spectroscopic Observations for SN\,2020cxd \label{tab:spec}}
\tablehead{
\colhead{Observation Date} & 
\colhead{Observation Date} & 
\colhead{Phase from explosion} &
\colhead{Telescope+Instrument} \\
\colhead{(YYYY MM DD)}  & 
\colhead{(JD)} &
\colhead{(Rest-frame days)} &
\colhead{} 
}
\startdata
2020 Feb 20 & 2458899.76 & 2.2 & LT+SPRAT \\
2020 Feb 20 & 2458899.94 & 2.4 & P60+SEDM \\
2020 Feb 25 & 2458904.99 & 7.5 & P60+SEDM \\
2020 May 22 & 2458991.94 & 94.4 & P60+SEDM \\
2020 Jun 25 & 2459026.48 & 129.0 & LT+SPRAT \\
2020 Jun 26 & 2459026.80 & 129.3 & P60+SEDM \\
2020 Jul 01 & 2459032.51 & 135.0 & NOT+ALFOSC \\
2020 Jul 28 & 2459058.86 & 161.4 & Gemini N+GMOS  \\
2020 Oct 19 & 2459141.74 & 244.2 & Keck1+LRIS  \\
\enddata
\end{deluxetable}

\begin{deluxetable}{lc}
\tablewidth{0pt}
\tabletypesize{\scriptsize}
\tablecaption{Measured Relative Galaxy Emission Line Fluxes
\label{tab:galflux}}
\tablehead{
\colhead{Line} &
\colhead{Flux (relative to H$\beta$)}
}
\startdata
{[}O II] $\lambda3727$ & $2.19 \pm 0.07$ \\
H$\gamma$ & $0.36 \pm 0.02$ \\
H$\beta$ & $1.00 \pm 0.02$\tablenotemark{a} \\
{[}O III] $\lambda4959$ & $0.74 \pm 0.03$ \\
{[}O III] $\lambda5007$ & $2.22 \pm 0.06$ \\
{[}N II] $\lambda6548$ & $0.14 \pm 0.01$ \\
H$\alpha$ & $4.08 \pm 0.09$ \\
{[}N II] $\lambda6583$ & $0.42 \pm 0.02$ \\
{[}S II] $\lambda6716$ & $0.40 \pm 0.01$ \\
{[}S II] $\lambda6731$ & $0.26 \pm 0.01$
\enddata
\tablenotetext{a}{Error bar indicates the relative flux error on the H$\beta$ measurement.}
\end{deluxetable}

\appendix 

\section{Diluted blackbody fit}
\label{sec:bb}

We fit a diluted blackbody (BB) function with multiple bands using the following formula:

\begin{equation}
    F_{\lambda}=(R/d)^2 \cdot \epsilon^2 \cdot \pi \cdot B(\lambda, T) \times 10^{-0.4 \cdot A_{\lambda}},
\end{equation}

where $F_{\lambda}$ is the flux at wavelength $\lambda$, $B$ is the Planck function, $A_{\lambda}$ is the extinction, $T$ is the temperature, $R$ is the radius, $d$ is the distance, and
$\epsilon$ is the dilution factor \citep{E96, Hamuy01, D05} that represents a general correction between the fitted BB distribution to the observed fluxes. We use the values from \citet{D05}. 

We also compare the absolute calibrated spectra of SN\,2020cxd to the $gri$ constructed BB functions. We integrated the de-reddened fluxes over the spectral bands using the trapezoidal rule, and calculated the ratio with the $gri$ inferred fluxes.
This method gives consistent results at early phases (ratio variance is small), 
but it does not work well in the nebular phase when the spectra become dominated by emission lines rather than the continuum.

\cite{Jager2020} compared fluxes inferred from diluted BB functions (from $VRI$) to those directly integrated including infrared (IR) data ($JHK$) for a sample of LL SNe II.
As shown in 
\citet[][their fig. 4,]{Jager2020} 
BB fitting with optical bands agrees well with the results from direct integration of the full dataset.
We also checked this with SN 2018hwm.
As shown in \citet[][their fig. 2]{Reguitti2020},
SN 2018hwm was routinely followed by ZTF in $g$ and $r$ until $\sim$ 60 rest frame days, and was thereafter observed in multiple bands up to 200 days.
We fit diluted BBs to the GP interpolated optical and IR
photometry of SN 2018hwm
every 2 days during this period. For each epoch, we calculated the ratio between the $gri$ inferred BB fluxes to those obtained by directly integrating the optical and $JHK$ spectral energy distributions (SED), and found the differences are always less than 10\%, which supports the validity of the \cite{Jager2020} approach.

\begin{figure*}
\centering
    \includegraphics[width=0.8\textwidth]{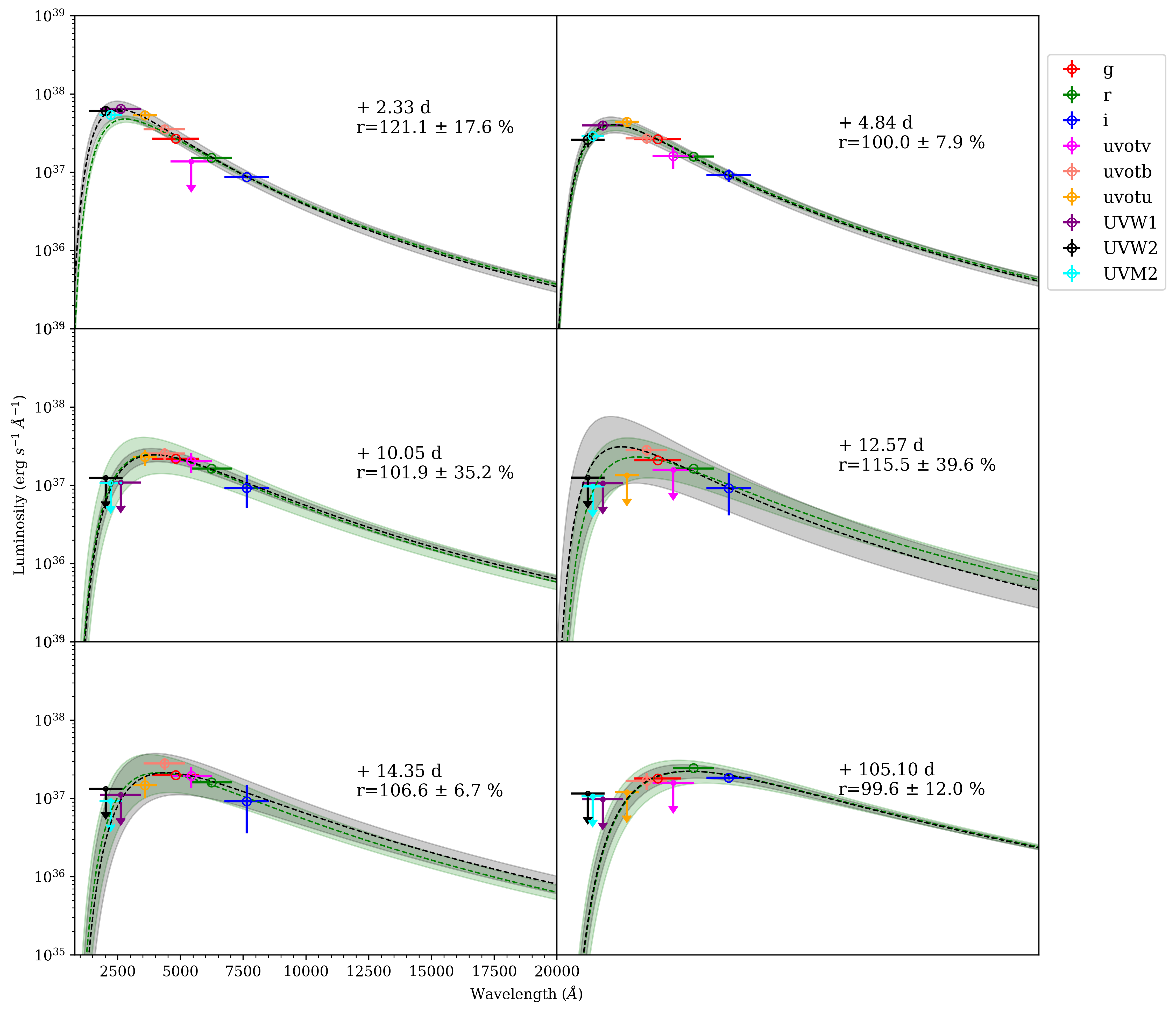}
    \caption{Blackbody fitting comparisons for SN\,2020cxd.
    The six subplots compare the $gri$+UV constructed BB at the six epochs when Swift UVOT data are available to the $gri$ inferred ones. Best fits of the full dataset are shown as the black dashed lines and the 3$\sigma$ uncertainties are overplotted as the grey shaded regions. Best fits and errors of the $gri$ dataset are instead shown as green dashed lines and shaded areas.
    The rest frame phase, as well as the ratio between the integrated luminosities of the full and $gri$ datasets, are provided to the right in each subplot.}
    \label{fig:uvfit}
\end{figure*}

In the UV region, \cite{Jager2020} extrapolate to 2000~\AA~from the $B$ and $V$ bands, assuming zero flux for even shorter wavelengths \citep{Lyman13}.
For SN\,2020cxd,  we have six epochs of Swift UVOT data, and can thus integrate the luminosity directly with the UV bands.
As shown in the six subplots  of Fig.~\ref{fig:uvfit}, we compare the $gri$ constructed BB to those inferred from using the full dataset including the Swift UV data.
The $gri$ inferred BB luminosities are similar to those obtained from the full dataset as well.

We adopt the BB fitting method to estimate the bolometric LC of SN\,2020cxd during its photospheric phase, and this is shown as a black dashed line in Fig.~\ref{fig:lumcompare}.
The dark orange shaded region in that figure represents the errors from the distance uncertainty, while the light orange area stands for the errors from the uncertainty of temperature of the fitted BB. For comparison, we also show the spectra integrated luminosity (green crosses) and $gri$+UV inferred luminosity (red crosses) in Fig. \ref{fig:lumcompare}.

\section{Bolometric correction method}
\label{sec:bc}

\begin{figure*}
\centering
    \includegraphics[width=0.8\textwidth]{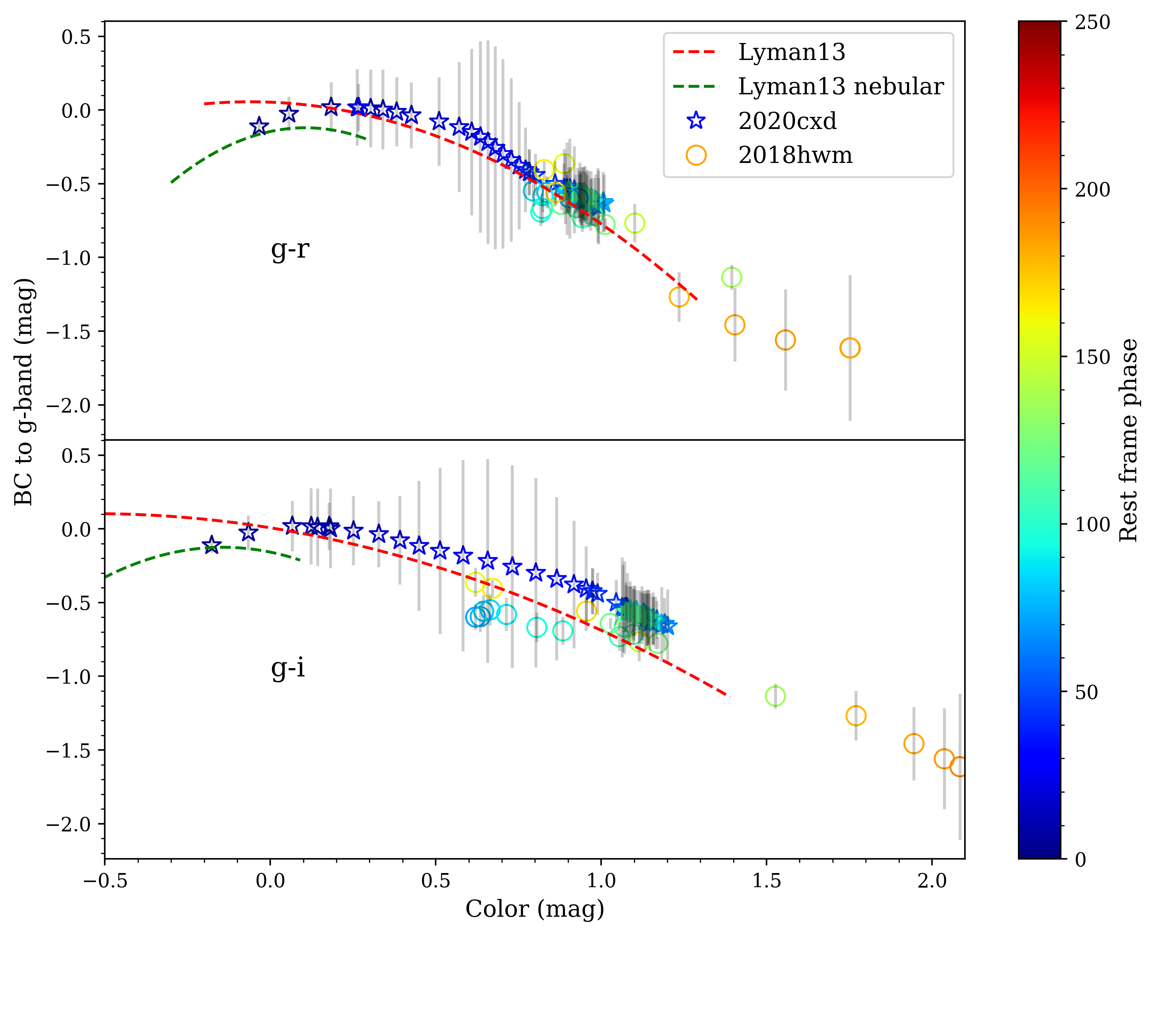}
    \caption{Bolometric correction of SN\,2020cxd in the $g$ band versus $g-r$ (upper panel) and $g-i$ (lower panel) colors. We have also included the corrections for SN 2018hvm as well as the comparison to the BC from \cite{Lyman13}. Overall, these are all similar which makes us confident in the applied corrections for SN\,2020cxd.}
    \label{fig:bc}
\end{figure*}

The BC is defined as:

\begin{equation}
    BC_{x}=M_{bol}-M_{x},
\end{equation}

where $M_{bol}$ is the bolometric magnitudes, and $M_{x}$ is the absolute magnitude of SN\,2020cxd in filter x.
\citet[][their fig. 5]{Jager2020} compare the $BC_{B}$ of the LL SN sample to show their similarity, which supports the validity of the empirical correlation found by \cite{Lyman13}.
We show the $BC_{g}$ evolution of SN\,2020cxd (as derived from our BB fits for the first 140 days) and of SN 2018hwm in Fig.~\ref{fig:bc}.
\cite{Lyman13} fitted $BC_{g}$ as a function of ($g-r$)/($g-i$) colors for a sample of Type II SNe, which is also shown as the red (for early phase) and green (for the nebular phase) dashed lines for comparison.
The $BC_{g}$ evolution of SN\,2020cxd is similar to that of SN 2018hwm, and also to that from the Lyman model during the optically thick phase.
Since the color of LL Type IIP SNe in the nebular phase is outside the fitting range of the Lyman model (the green dashed line),
we use the $BC_{g}$ of SN 2018hwm to estimate the bolometric LC of SN\,2020cxd in the nebular phase, shown as the blue dashed line in 
Fig.~\ref{fig:lumcompare}. In fact, it is the $BC_{r}$ for SN 2018hwm that we use for SN\,2020cxd, since we do not have $g$-band data on the tail. 
The dark blue shaded region corresponds to the errors from the distance uncertainty of SN\,2020cxd, while the light blue shaded area marks the errors from the BB fitting and validity.

\section{Monto carlo fitting code with semi-analytic models}
\label{sec:mc}

This code was first developed by \cite{Nagy16}, which
generate modelled LCs of CC SNe for a variety of parameters e.g. the ejected mass, the initial progenitor radius, the total explosion energy, and the synthesized nickel mass \citep[following early work of ][]{Arnett1989, Popov, Blinnikov, Nagy14}, and was further developed by \cite{Jager2020} who added a markov chain monte carlo (MCMC) method.
The model is based on a two-component configuration consisting of a uniform dense stellar core and an extended low-mass envelope where the density decreases as an exponential function. 
After comparison, \citet{Nagy16} conclude that the results from the two-component semi-analytic LC model are consistent with current state-of-the-art calculations for Type II SNe, so the estimated fitting parameters can be used for preliminary studies awaiting more sophisticated hydrodynamic modelling.

\begin{figure*}
\centering
    \includegraphics[width=0.8\textwidth]{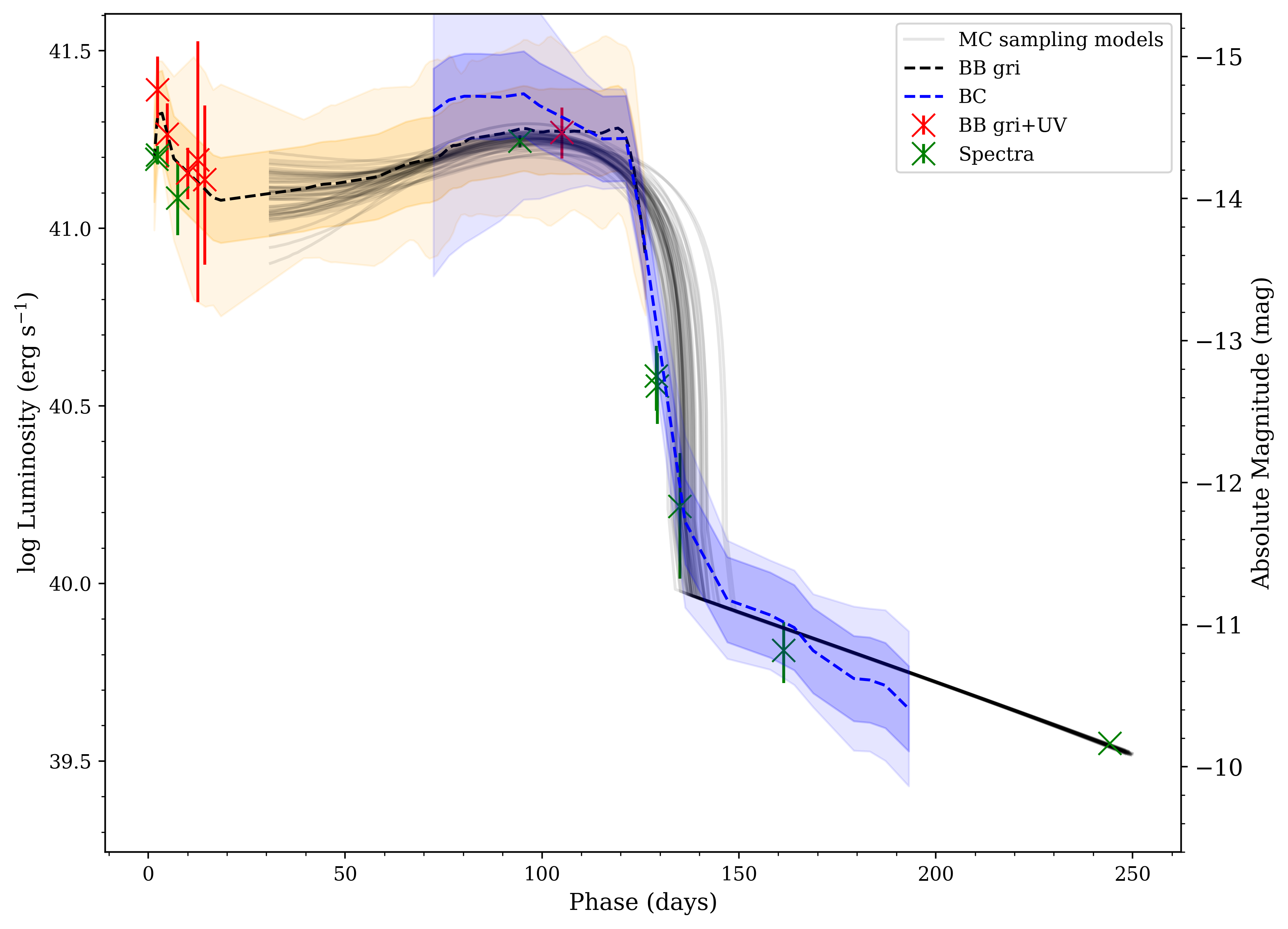}
    \caption{SN\,2020cxd LC comparison between observations and sampled MCMC fits (solid black lines). 
    The black dashed line represents the observed LC of SN\,2020cxd at early phases from BB fitting, with the orange shaded region as errors from the distance uncertainty (dark) and from the fitted BB (light). 
    The blue dashed line is the observed LC of SN\,2020cxd at later epochs from the BC approach, with the blue shaded region as errors from the distance uncertainty (dark) and the fitted BB (light).
    The green crosses stands for the epochs with fluxes from direct spectral integration, while the red crosses is from BB fits with $gri$+UV photometry. }
    \label{fig:lumcompare}
\end{figure*}

We also compared the semi-analytic fitting results from \cite{Jager2020} to hydrodynamical modelling results from \cite{Martinez2020}, 
and found that the estimated ejecta masses are similar. 
For SN 2005cs, the best-fit ejecta mass with the analytic model is 8.84 \Msun, which is similar to the hydrodynamical result of 8 \Msun. 
For SN 2004et, the estimated ejecta masses of the semi-analytic and hydrodynamical models are both 13 \Msun.
As an additional test, we fit SN 2018hwm with the semi-analytic LC fitting code, and obtain $7.6\pm1.3$ \Msun\ for the ejecta mass, which is consistent with the hydrodynamic modelling result of \cite{Reguitti2020} of 8 \Msun. These comparisons provide some confidence in the derived ejecta masses from the simple analytical fits provided here.
The fits to the LC of SN 2020cxd are shown in Fig.~\ref{fig:lumcompare}.

\end{document}